  \documentclass[12pt]{article}
  \usepackage{latexsym}
  \usepackage{epsf}
  
  \makeatletter
  \@addtoreset{equation}{section}
  \makeatother
  
  \newcommand{\bra}[1]{\langle{#1}|}
  \newcommand{\ket}[1]{|{#1}\rangle}
  \def\one{{\hbox{1\kern-.8mm l}}}
  \def\be{\begin{equation}}
  \def\ee{\end{equation}}
  \def\ba{\begin{array}}
  \def\ea{\end{array}}
  \def\bea{\begin{eqnarray}}
  \def\eea{\end{eqnarray}}
  
  \def\orbifold{$(-1)^{F_L} \cdot {\cal I}_4~$}
  \def\orientifold{$\Omega \cdot {\cal I}_4~$}
  \def\Z {{\rm Z\!\!Z}}

\begin{document}
  \begin{titlepage}
  \pagestyle{plain}
  \begin{flushright}
  \footnotesize
  DAMTP-2000-20\\
  IC-2000-13\\
  MRI-PHY/P20000201\\
  {\bf hep-th/0003033}\\
  \normalsize
  \end{flushright}

  \vskip 30pt

  \begin{center}
  {\LARGE \bf The Spacetime Life}\\
  \vskip 7pt
  {\LARGE \bf of a non-BPS D-particle}
  \vskip 25pt

  {\large Eduardo Eyras$^{~a}$ and Sudhakar Panda$^{~b,c}$}

  \vskip 28pt

  ${}^a$
  {\em Department of Applied Mathematics and
  Theoretical Physics,}\\
  {\em University of Cambridge,}\\
  {\em Wilberforce Road, CB3 0WA Cambridge, U.K.}\\
  {\tt E.Eyras@damtp.cam.ac.uk}

  \vskip 12pt

  ${}^b$
  {\em The Abdus Salam International Centre for Theoretical Physics}\\
  {\em Strada Costiera 11, I-34014, Trieste, Italy}\\

  \vskip 12pt

  ${}^c$
  {\em Mehta Research Institute of Mathematics and Mathematical Physics}\\
  {\em Chhatnag Road, Jhoosi}\\
  {\em Allahabad 211019, India}\\
  {\tt panda@mri.ernet.in}

  \vskip 22pt


  {\bf Abstract}
  \end{center}

  \begin{quotation}
  \small
  We investigate the classical geometry
  generated by a stable non-BPS D-particle.
  We consider the boundary state of
  a stable non-BPS D-particle in the covariant formalism
  in the type IIB theory orbifolded by
  \orbifold.
We calculate the scattering amplitude between two D-particles
in the non-compact and compact orbifold and analyse the short and long
distance behaviour. At short distances we find no 
force at order $v^2$ for any radius,
and at the critical radius we find a BPS-like behaviour up
to $v^4$ corrections for long and short distances. 
  Projecting the boundary state on the massless states
  of the orbifold closed string spectrum we obtain
  the large distance behaviour of the classical solution
  describing this non-BPS D-particle in the non-compact and compact cases.
  By using the non-BPS D-particle
  as a probe of the background geometry
  of another non-BPS D-particle, we recover the no-force condition
  at the critical radius  and the $v^2$ behaviour of the probe. 
Moreover, assuming that the no-force
  persists for the complete geometry
  we derive part of the classical
  solution for the non-BPS D-particle.

  \end{quotation}
  \end{titlepage}

  \newpage

  \tableofcontents

  \section{Introduction}

  BPS D-branes enjoy a double life. On the one hand as a
  conformal field theory described by open
  strings with Dirichlet boundary conditions
  \cite{D-branes}, and on the other hand
  as classical solitons of supergravity\footnote{See \cite{p-branes,solitons}
  and references therein.}.
  For a single D-brane the regimes of validity of these two
  descriptions are complementary. The conformal field theory description
  is valid at weak coupling; whereas the supergravity solution
  corresponds to a strongly interacting gravitational system, which
  corresponds to strong coupling.
  On the other hand, for a large number of branes $N$ the system
  can be well described by a classical solution
  at weak coupling \cite{Maldacena-Strominger}.
The two main properties of the BPS D-branes
  that assure this consistent {\it dual} behaviour
  are the fact that BPS-branes preserve
  a fraction of supersymmetry, hence they are stable under
  variations of the string coupling; and the
  fact that one can consider a superposition of a large number of D-branes
  in such a way that they still preserve a fraction of supersymmetry.
  These properties make possible, for instance, the
  entropy calculation of black-holes using D-branes
  \cite{entropy}.
  On the other hand, D-branes can be studied in more general situations,
  for which spacetime supersymmetry is not preserved.
  This is the case of the non-BPS D-branes
  \cite{Sen-0,Sen-1,Bergman-Gaberdiel-1,Sen-2,
  Witten-Ktheory,Sen-5,Sen-review,Lerda-review,Gaberdiel-review}.
  These branes have an exact 
  conformal field theory description, whose consistency conditions do not rely
  on spacetime supersymmetry
  \cite{Sen-1,Bergman-Gaberdiel-1,Sen-2,Sen-5}.
  Having such a precise description for the non-BPS D-branes
  in the conformal field theory side, it is natural to wonder
  whether the non-BPS D-branes also enjoy a double life
  and have a description in terms of a classical solution of
  a certain effective (super)gravity theory.

  It has been recently suggested
  in \cite{Harvey-etal}
  that the unstable non-BPS branes of type II theories
  might have a gravitational counterpart
  described by a gravitational sphaleron\footnote{See \cite{Volkov} and
  references therein.}.
  These are unstable solutions with finite energy
  which interpolate between two
  (possibly distinguishable) vacuum configurations.
  However, these solutions are unstable
  and probably do not remain valid classically.
  On the other hand, we expect that consistent classical solutions
  will be related to non-BPS branes which have some
  stability properties similar to the BPS D-branes.

  In order to find the appropriate conditions
  for constructing a classical solution for a non-BPS D-brane we must
  find out which properties of the D-branes are such that (1) they assure the
  validity of a classical supergravity description
  and (2) they are not necessarily related to fractional supersymmetry.
  These properties certainly comprise {\it stability},
  which ensures that the state survives at strong coupling. Moreover,
  stability
  can be based on grounds different from supersymmetry, for instance, being
  the
  lightest object carrying a certain quantum number \cite{Sen-0,Sen-1}.
  Another key property is the fact that we can superpose
  an arbitrary number of parallel D-branes, which is the same as having
  a
  {\it no-force} condition at all distances \cite{solitons,Lifschytz,Tseytlin}.
  This is a consequence
  of the BPS property, which
  can exist independently of supersymmetry.
  Therefore, it seems natural that
  in order to find a classical
  solution for a non-BPS D-brane, this brane should be {\it stable}
  and enjoy a {\it no-force} property.
  Stable non-BPS D-branes have been found in different theories
  \cite{Sen-1,Bergman-Gaberdiel-1,Sen-2,Witten-Ktheory,Sen-5,
  Bergman-Gaberdiel-2,Gaberdiel-Stefanski,Sagnotti}. However,
  only non-BPS D-branes in a certain orbifold
  of type II theories 
are known to enjoy both stability and the no-force property
  \cite{Gaberdiel-Sen}. There it was shown that at a particular {\it critical}
  radius
  of the compact orbifold the non-BPS branes develop a
  Bose-Fermi degeneracy at one loop. The critical radius
  is in fact the value beyond which the non-BPS D-brane becomes
  unstable and can decay into a pair of BPS D-branes.

  In this article we study the particular case of a
  stable non-BPS D-particle in type IIB string
  theory\footnote{The gravity duals of BPS branes in orbifolds
  have been considered before in \cite{Klebanov-etal}.} orbifolded
  by \orbifold . This D-particle is a truncated D-brane 
\cite{Gaberdiel-Stefanski}, is charged electrically
  under a twisted R-R 1-form field
  and it is the strong-weak coupling dual of a non-BPS state
  in the orientifold \orientifold  of type IIB,
  charged under the $U(1)$ field of the D5-O5 system \cite{Sen-1}.
  The coupling of the non-BPS D-particle to a
  twisted R-R vector field is the origin of the stability of
  this non-BPS D-particle \cite{Bergman-Gaberdiel-1}.
  Moreover, at a critical radius of the compact orbifold
  this D-particle meets all the requirements suggesting the existence of
  a classical solution.
  We use the technique of the boundary state in the covariant
  formalism\footnote{For a recent review see
  \cite{Divecchia-Liccardo}.} \cite{covariant-state}
  to describe the non-BPS D-particle.
We analyse the interaction potential between two non-BPS D-particles
in relative motion. Remarkably, we find no force at order $v^2$ as for BPS 
D-branes. At the critical radius, the static force is moreover vanishing,
hence the non-BPS D-particle presents a BPS-like behaviour, up to
$v^4$ corrections. Unlike for BPS branes, we find no matching between 
the $v^4$ terms in the open and closed string description.

  Using the boundary state
  for BPS D-branes,
  the long distance behaviour of the classical
  massless fields generated by the D-brane was computed in\footnote{The same 
results were obtained earlier in \cite{Garousi-Myers} 
using different techniques.}
  \cite{Divecchia-1,Divecchia-3}, 
  and it was shown that
  the asymptotic behaviour of the corresponding classical solution
  is precisely recovered.
  A BPS D$p$-brane is described by a boundary state with two parts,
  the NS-NS part and the R-R part. They generate 
the asymptotic behaviour of NS-NS
  massless fields (metric and dilaton), and
  of the R-R massless fields (R-R $(p+1)$-form potential), respectively.
  In this paper we implement the same technique to obtain the long distance
  behaviour of the non-BPS D-particle geometry.
  This is given by the asymptotic form of a metric and a dilaton propagating
  in the bulk, and a twisted R-R 1-form potential propagating in the
  orbifold fixed plane.

  One difficulty about recovering the full solution
  for the non-BPS D-particle from its asymptotic
  form is that, in principle, there might be many different
  possible geometries with the same asymptotic structure.
  In order to restrict these geometries we will assume that the no force
condition also takes place when one consider the complete geometry
at the critical radius, as it happens for BPS branes \cite{Tseytlin}.
That is, we take into account the extra pieces 
that one would need to add 
to the asymptotic behaviour in order to recover the full form of metric.
On the other hand, although the Bose-Fermi degeneracy occurs
at any distance between the non-BPS D-branes 
\cite{Gaberdiel-Sen}, at short distances, open strings loops might
spoil this property. In fact, it has been recently proposed in
\cite{Neil} that even at
1-loop in the open string theory, the no-force is removed in favour of 
another vacuum configuration in which the branes attract each other.
For these reasons, our assumptions will be only acceptable for distances
much larger than the string scale, which is also the range
of validity of the classical solution for a BPS D-brane
\cite{DKPS}.

Using the non-BPS D-particle as a probe
in the background of another non-BPS D-particle we recover
the no-force behaviour at the critical radius.
Moreover, under the assumptions presented above we are able to
obtain part of the complete metric, dilaton and twisted R-R 1-form
generated by the D-particle.
The velocity dependent part of the
brane action multiplies a flat metric,
agreeing with the vanishing of the $v^2$ in the
interaction potential. 
Extending this property for the complete geometry of the 
non-BPS D-particle source, 
we are able to find more properties of the classical geometry.
  We expect that the solution is consistent classically
  at the critical radius only, since it is only there where one can consider
  a superposition of a large number of D-particles $N$.

  We find a diagonal metric with $SO(5)\times SO(4)$ symmetry
  (in Einstein frame):
  \begin{displaymath}
  ds_{10,E}^2 = g_{00}(y) dt^2 + g_{mn}(y) dy^m dy^n + g_{ij}(y) dx^i dx^j \,
  ,
  \end{displaymath}
  where we do the split $\mu = (0,m,i)$ according to the orbifold symmetry,
  and ${\cal I}_4$ acts on the $i$ directions.
  By using the assumptions mentioned above we are able to find the form of
  two of the components of the metric at the critical radius:
  \bea
  g_{00} (y) &=& - \, \left( 1 + {\kappa_6 T_0 \over 2a \Omega_4}
  (2\pi^2 \alpha^\prime)^{-1} \, {1 \over |y|^3}
  + \dots \right)^{-{7\over 6}a}
  \, ,\nonumber \\
  g_{mn} (y) &=& \left( 1 + {\kappa_6 T_0 \over 2a \Omega_4}
  (2\pi^2 \alpha^\prime)^{-1} \, {1 \over |y|^3}
  + \dots \right)^{{1\over 6}a} \delta_{mn} \, .\nonumber
  \eea
  The expressions are given in terms of a parameter $a$, and some
  possible extra dependence in $|y|^n$, $n < -3$
  denoted by $\dots$ which remain to be
  determined. We find no expression for $g_{ij}$ since, as will be explained
  in Section 4, the
  D-particle cannot probe the precise geometry in these directions.
  Moreover, the form of the dilaton and twisted R-R potential
  at the critical radius are found to be:
  \bea
  {\rm e}^\phi &=& \left( 1  + {\kappa_6 T_0 \over 2a \Omega_4}
  (2\pi^2 \alpha^\prime)^{-1} \, {1 \over |y|^3} + \dots 
  \right)^a
  \, ,\nonumber \\
  {\cal C}^{(1)}_0 &=& \left( 1 + {\kappa_6 Q_0 \over 4 a \Omega_4}
  \, {1 \over |y|^3} + \dots \right)^{-{4\over 3}a} -1
  \, .\nonumber
  \eea
Here $T_0$ is the tension of the D-particle and is related to its
charge $Q_0$ by $T_0 = Q_0 \pi^2 \alpha^\prime$, hence at the critical radius
the fields above are given in terms of a single function.

  This article is organised as follows.
  In Section 2 we carry out the construction of the covariant boundary state
  for the non-BPS D-particle in the orbifold of type IIB
  generated by \orbifold, for the non-compact and compact cases.
We also evaluate the amplitude for D-particles in relative motion and analyse
the long and short distance behaviour of the interaction.  
In Section 3 we evaluate the asymptotic behaviour of the massless
  fields excited by the non-BPS D-particle, for the non-compact and
  compact cases. In Section 4 we recover the no-force property at the critical
  radius and derive part of the complete classical solution, by using the
  assumptions mentioned above.
  For completeness, we also include the derivation of the
  zero-mode of the twisted R-R boundary state in Appendix A, and
  the explicit form of the complete covariant boundary state for the D-particle
  in Appendix B.


  \section{Stable non-BPS D-particle in IIB/$(-1)^{F_L} {\cal I}_4$}

  The stable non-BPS D-particle of type IIB string theory orbifolded by
  $(-1)^{F_L} \cdot {\cal I}_4$ was described in
  \cite{Sen-0,Sen-1,Bergman-Gaberdiel-1}.
  It corresponds to the S-dual of a massive particle-like state
  in the system of a D5-brane on top of an orientifold 5-plane,
  which is stable but non-BPS \cite{Sen-0}.
  Its stability is due to the fact that
  this particle is the lightest state charged under the $U(1)$ gauge field
  of the D5-O5 worldvolume theory. Similarly, the stability of the
  D-particle is due to its charge under a twisted R-R 1-form,
  which can be
  identified with the $U(1)$ gauge field in the worldvolume of a NS-5
  brane on top of the orbifold fixed plane.

  Let us consider first the non-compact theory, i.e. type IIB
  in Minkowski space orbifolded by \orbifold.
  The operator ${\cal I}_4$
  corresponds to a reflection in the directions
  $x^i$, $i=6,7,8,9$.
  In the non-compact case, the orbifold contains a fixed
  plane at $x^6=x^7=x^8=x^9=0$, hence
  the orbifold breaks the $SO(1,9)$ symmetry down
  to $SO(1,5) \times SO(4)$.
  The operator $F_L$ is the spacetime fermion number of the left-sector,
  hence $(-1)^{F_L}$ changes the sign of the of the
  R-R groundstate, without having any action on the oscillators.
  The closed string spectrum of the orbifold theory consists of an untwisted
  sector,
  given by type IIB
  states which are invariant under \orbifold, and a twisted sector
  localised on the orbifold fixed plane.
  If we split the coordinates as $X^\mu=(X^\alpha,X^i)$,
  where $X^\alpha$, $\alpha=0,\dots,5$,
  is longitudinal to the fixed plane, and $X^i$, $i=6,\dots,9$, transverse,
  the oscillators of the twisted sector have the following modding:
  \begin{eqnarray}
  \mbox{twisted NS} : &
  \left\{
  \begin{array}{ll}
  \alpha^\alpha_n \, ,& n \in \Z \\
  \alpha^i_r \, ,& r \in \Z + 1/2
  \end{array}\right.
  \qquad
  \left\{
  \begin{array}{ll}
  \psi_r^\alpha \, ,& r \in \Z + 1/2 \\
  \psi_n^i \, ,& n \in \Z
  \end{array}\right. \nonumber \\
  \mbox{twisted R} : &
  \left\{
  \begin{array}{ll}
  \alpha^\alpha_n \, ,& n \in \Z \\
  \alpha^i_r \, ,& r \in \Z + 1/2
  \end{array}\right.
  \qquad
  \left\{
  \begin{array}{ll}
  \psi_n^\alpha \, ,& n \in \Z \\
  \psi_r^i \, ,& r \in \Z + 1/2
  \end{array}\right.
  \label{twisted-modding}
  \end{eqnarray}
  There are 4 fermionic zero-modes in the NS sector
  in the directions $i=6,7,8,9$, and 6 fermionic zero-modes in the R-sector
  in the directions $\alpha=0,1,\dots,5$.
  Since the intercept vanish for both sectors,
  the twisted groundstate will be given by these zero-modes.
  The twisted NS-NS sector gives a vector of $SO(4)$, or equivalently
  4 scalars under $SO(1,5)$. On the other hand, the
  twisted R-R sector gives rise to a
  vector under $SO(1,5)$.
  From the point of view of the fixed plane, the supersymmetries are generated
  by
  two Weyl supercharges of different chirality. This corresponds
  to $(1,1)$ supersymmetry in 6 dimensions.
  In fact, the massless twisted sector can be identified
  with the worldvolume degrees of freedom of a NS-5 brane of type IIB
  \cite{Sen-1,Kutasov,Majumder-Sen}.



  \subsection{The D-particle Boundary State}

  In this section we construct the boundary state for
  the stable non-BPS D-particle in type IIB orbifolded
  by \orbifold.
  This has been carried out in the light-cone gauge in
  \cite{Bergman-Gaberdiel-1}.
  Here we use the covariant formalism
  for the boundary state, which has not been implemented before
  for non-BPS D-branes in orbifolds.
  The D-particle boundary state is made up
  of an untwisted NS-NS part and a twisted
  R-R part\footnote{We use the subindex NS and R in the boundary states for
  short.} \cite{Bergman-Gaberdiel-1}:
  \be
  | D0 {\cal i} = | D0 {\cal i}_{\rm NS,U} + |D0 {\cal i}_{\rm R,T} \, .
  \ee
  The NS-NS boundary state is defined as the GSO-invariant combination
  of boundary states $\ket{D0,\eta}_{\rm NS,U}$, with $\eta=\pm 1$,
  which turns out to be \cite{Bergman-Gaberdiel-1}:
  \be
  |D0 {\cal i}_{\rm NS,U} =
  {\cal P}_{\rm GSO,U} \ket{D0,+}_{\rm NS,U} =
  {1 \over 2}
  \left( | D0,+ {\cal i}_{\rm NS,U} - |D0,- {\cal i}_{\rm NS,U} \right) \, ,
  \ee
  where the GSO projector in the NS-NS sector is given by
  \be
  {\cal P}_{\rm GSO,U} = {1 \over 4}\left(1 - (-1)^{F+G}\right)
  \left(1 - (-1)^{\tilde{F} + \tilde{G}} \right) \, ,
  \ee
  with $F$ and $G$ the (worldsheet) fermion and superghost number
  operators, respectively:
  \be
  F = \sum_{m=1/2}^{\infty} \psi_{-m} \cdot \psi_m \, ,\qquad
  G = - \sum_{m=1/2}^{\infty}
  \left( \gamma_{-m} \beta_m + \beta_{-m} \gamma_m
  \right) \, ,
  \ee
  and $\beta$, $\gamma$ are the superghosts. For the right movers these
  operators are analogously defined.
  The state $\ket{D0,\eta}_{\rm NS,U}$
  takes the form \cite{Divecchia-2}:
  \be
  \ket{D0,\eta}_{\rm NS,U} = {T_0 \over 2}
  \ket{D0_X}\, \ket{D0_{\rm gh}}\,{\ket{D0_\psi,\eta}}_{\rm NS}
  \,{\ket{D0_{\rm sgh},\eta}}_{\rm NS} \, ,
  \ee
  where $T_0$, a constant related to the tension of the D-particle, is to
  be
  determined later.
  There is a bosonic and a fermionic part,
  ($\ket{D0_X}$ and $\ket{D0_\psi,\eta}_{\rm NS}$),
  and also a ghost and a superghost part
  ($\ket{D0_{\rm gh}}$ and $\ket{D0_{\rm sgh},\eta}_{\rm NS}$).
  The boundary state $\ket{D0,\eta}_{\rm NS,U}$
  is very similar to the NS-NS boundary state for type II branes.
  In fact, the only piece modified
  by the orbifold with respect to the
  type II case is the zero-mode part of $\ket{D0_X}$.
  For the bosonic untwisted boundary state 
  the most general conditions invariant
  under the orbifold symmetry are
  \bea
  \partial_\tau X^0|_{\tau=0} \ket{D0_X} &=& 0 \, ,\nonumber \\
  X^p |_{\tau=0} \ket{D0_X} &=& y^p \, ,\qquad p=1,\dots , 5 \, ,\\
  X^i |_{\tau=0} \ket{D0_X} &=& 0 \, ,\qquad i=6,\dots , 9 \, ,\nonumber
  \eea
  from which we can deduce that
  \be
  \ket{D0_X} = \delta^{(5)}(\hat{q}^p - y^p) \, \delta^{(4)}(\hat{q}^i) \,
  {\rm exp}\biggl[-\sum_{n=1}^\infty \frac{1}{n}\,
  \alpha_{-n}\cdot S\cdot
  \tilde{\alpha}_{-n}\biggr]\,
  \ket{k=0} \, .
  \ee
  Note that the 
  D-particle position in the directions $i=6,\dots,9$ is restricted
  by the orbifold symmetry to be
  on the fixed plane.
  The other pieces of the untwisted
  boundary state are not modified by the orbifold and
  are the same as for the type II case.
  These are explicitly given in Appendix \ref{appendixB}.
  Moreover, since there are nine Dirichlet directions for the case of the
  D-particle we
  have\footnote{We take the Minkowski metric to be mostly plus.}
  $S_{\mu\nu}= - \delta_{\mu\nu}$.

  The twisted R-R boundary state that we denote by
  $\ket{D0}_{\rm R,T}$ is defined as the GSO invariant combination
  of boundary states $\ket{D0,\eta}_{\rm R,T}$ and
  is given by \cite{Bergman-Gaberdiel-1}:
  \be
  \ket{D0}_{\rm R,T} = {\cal P}_{\rm GSO,T} \ket{D0,+}_{\rm R,T}
  = {1\over 2}\left( \ket{D0,+}_{\rm R,T} + \ket{D0,-}_{\rm R,T} \right) \, .
  \ee
  In this sector the GSO-operator is given by:
  \be
  {\cal P}_{\rm GSO,T} = {1 \over 4}\left(1 + (-1)^{F+G}\right)
  \left(1 - (-1)^{\tilde{F} + \tilde{G}} \right) \, ,
  \ee
  with $F$ and $G$ the (worldsheet) fermion and superghost number
  operators in the twisted R-sector, respectively:
  \bea
  (-1)^F &=& \Psi \, (-1)^{\sum\limits_{m=1}^{\infty} \psi^\alpha_{-m}
  \eta_{\alpha\beta} \psi^\beta_m}
  \, (-1)^{\sum\limits_{r=1/2}^{\infty} \psi^i_{-r}
  \delta_{ij} \psi^j_r}
  \, ,\nonumber \\
  G &=& -\gamma_0 \beta_0 - \sum_{m=1}^{\infty}
  \left( \gamma_{-m} \beta_m + \beta_{-m} \gamma_m
  \right) \, ,
  \label{RR-GSO-operators}
  \eea
  and similarly for the right movers.
  Here $\Psi$ (and $\tilde\Psi$) represent the zero-mode parts of the
  GSO-projectors, which are explicitly given in Appendix \ref{appendixA}.
  The twisted R-R boundary state $\ket{D0,\eta}_{\rm R,T}$
  is given by:
  \be
  \ket{D0,\eta}_{\rm R,T} = {Q_0 \over 2}
  \ket{D0_X}_{\rm T} \ket{D0_{\rm gh}}
  \ket{D0_\psi, \eta}_{\rm R,T} \ket{D0_{\rm sgh}, \eta}_{\rm R} \, ,
  \label{RR-twisted-state}
  \ee
  where $Q_0$ is a normalisation factor related to the charge density of
  the brane and to be determined below. Notice that since this D-particle
  is non-BPS, $Q_0 \neq T_0$, unlike the BPS D-branes.
  The (super)ghosts are not affected by the orbifold
  \cite{orbifold-vertices}, hence the corresponding pieces
  have the same form as for a type II R-R boundary state.
  Since in the twisted R-R sector the bosons have integer modding along the
  orbifold fixed plane and half-integer modding along the orbifolded
  directions,
  the state (\ref{RR-twisted-state}) have zero-modes
  along the fixed plane only. Accordingly,
  the twisted bosonic part takes the form:
  \be
  \ket{D0_X}_{\rm T} = \delta^{(5)}(\hat{q}^p - y^p) \,
  {\rm exp}\biggl[\sum_{n=1}^\infty \frac{1}{n}\,
  \alpha^\alpha_{-n} \delta_{\alpha\beta} \tilde{\alpha}^\beta_{-n}
  + \sum_{n=1/2}^\infty \frac{1}{n}\,
  \alpha^i_{-n} \delta_{ij} \tilde{\alpha}^j_{-n}
  \biggr]\,
  \ket{k=0} \, .
  \ee
  The fermions in the twisted R-R sector have integer
  modding along the fixed plane and half-integer modding along the
  orbifolded directions. As a consequence (\ref{RR-twisted-state})
  have fermionic
  zero-modes on the fixed plane only.
  The fermionic overlap conditions now read:
  \bea
  \left(\psi_m^{\alpha} - i \eta \, S^\alpha{}_\beta
  \, \tilde{\psi}_{-m}^\beta \right)
  \ket{D0_{\psi}, \eta}_{\rm R,T} &=& 0 \, ,\qquad m \in \Z \, ,\nonumber \\
  \left( \psi_r^{i} - i \eta \, S^i{}_j \, \tilde{\psi}_{-r}^j \right)
  \ket{D0_{\psi}, \eta}_{\rm R,T} &=& 0 \, ,\qquad r \in \Z+{1\over 2} \, .
  \label{twisted-overlap}
  \eea
  From them we can deduce:
  \be
  \ket{D0_\psi,\eta}_{\rm R,T} = - {\rm exp}
  \biggl[- {\rm i}\eta\sum_{m=1}^\infty
  \psi^\alpha_{-m} \delta_{\alpha \beta} \, \tilde{\psi}^\beta_{-m}
  - {\rm i}\eta\sum_{m=1/2}^\infty
  \psi^i_{-m} \delta_{ij} \, \tilde{\psi}^j_{-m}\biggr]
  \,\ket{D0,\eta}^{(0)}_{\rm R,T} \, .
  \ee
  The zero-mode part $\ket{D0,\eta}^{(0)}_{\rm R,T}$ is
  determined from the zero mode overlap conditions
  in (\ref{twisted-overlap}) and is
  constructed upon the twisted R-R groundstate:
  \be
  \ket{D0,\eta}^{(0)}_{\rm R,T} = M_{ab} \ket{a}_{\rm T}
  \widetilde{\ket{b}}_{\rm T} \, ,
  \ee
  where $a$ and $b$ are spinor indices of $SO(1,5)$.
  The explicit form of $M_{ab}$ and its derivation
  are given in Appendix \ref{appendixA}.

  Having constructed the covariant boundary state
  for the non-BPS D-particle, we determine next
  its tension and charge using the open-closed
  string consistency condition for boundary states.
  The interaction between two D-particles separated by a distance
  $y$ in the fixed plane is given by the amplitude between two
  boundary states
  located at a relative distance $y$ with respect to each other,
  with the insertion of a closed string propagator
  \be
  \bra{D0} {\cal D} \ket{D0} \, ,
  \ee
  with
  \be
  {\cal D}_{a} = {\alpha^\prime \over 4\pi} \int_{|z| \leq 1}
  {d^2 z \over |z|^2} z^{L_0 -a} \bar{z}^{\tilde{L}_0 - a} \, ,
  \ee
  where $a=1/2$ in the untwisted NS-NS sector and
  $a=0$ in the twisted R-R sector. Moreover,
  one needs to take the appropriate
  moddings in the expression for $L_0$ and $\tilde{L}_0$ in the twisted
  sector. Also care must be taken in computing the matrix elements involving
  the
  zero modes of the superghosts and fermions in the twisted R-R sector. This
  is because, in general, the superghost zero modes produce infinite number of
  terms with any superghost number contribution, hence a
  regularisation is needed. We use the same regularisation 
  as in   \cite{Divecchia-2}.
  Defining $\ket{D0,\eta}^{(0)}_{\rm {R,T}} = \ket{D0_\psi
  ,\eta}^{(0)}_{\rm {R,T}}~
  \ket{D0_{\rm sgh},\eta}^{(0)}_{\rm R}$ we give below the regularised result:
  \be
  {}^{(0)}_{\rm {R,T}}\bra{D0,\eta_1} {D0,\eta_2}\rangle^{(0)}_{\rm {R,T}}
  \, = \lim\limits_{\rho \to 1} \,
  {}^{(0)}_{\rm {R,T}}\bra{D0,\eta_1}
  \rho^{2F_0 + 2G_0}
  \ket{D0,\eta_2}^{(0)}_{\rm {R,T}}
  = ~ - 4
  \delta_{\eta_1\eta_2, 1} \, ,
  \ee
  where $F_0$ and $G_0$ are the zero-mode parts of the
  operators $F$ and $G$, implicitly given in
  (\ref{RR-GSO-operators}).
  Making a change of variables according to
  $|z|= {\rm e}^{-\pi\ell}$
  and $d^2z = -\pi {\rm e}^{-2\pi \ell} d \ell d\phi$ we
  find:
  \bea
  \bra{D0} {\cal D} \ket{D0} &=&
  {V_1 \alpha^\prime \pi \over 16} (2\pi^2 \alpha^\prime)^{-9/2}
  \int_0^\infty d \ell \, \ell^{-9/2} \,
  {\rm e}^{-{ y^2 \over 2\pi \alpha^\prime \ell}}
  \, \times
  \label{amplitude-non-compact-v=0}
\\
  &&\hspace{-2.5cm}\times \left\{
  ({T_0})^2 \,
  {f_3^8 ({\rm e}^{-\pi\ell}) - f_4^8 ({\rm e}^{-\pi\ell}) \over
  f_1^8 ({\rm e}^{-\pi\ell})} -
  ({Q_0})^2 (2\pi^2 \alpha^\prime \ell)^2 \,
  {f_2^4 ({\rm e}^{-\pi\ell}) f_3^4 ({\rm e}^{-\pi\ell}) \over
  f_1^4 ({\rm e}^{-\pi\ell}) f_4^4 ({\rm e}^{-\pi\ell})} \right\} \, ,
  \nonumber \\
  \nonumber
  \eea
  where $V_1$ is the (infinite) length of the D-particle worldline and $f_i$
  are
  functions of ${\rm e}^{-\pi\ell}$ defined in the usual manner.
  We can make a worldsheet transformation, $\ell = {1/\tau}$, to express the
  above
  closed string channel result in the open string channel. Note that we have
  considered both closed string and open string to have same periodicity in
  the
  spatial direction of the worldsheet.
  Open-closed string consistency then requires
  that the result must be equal to the
  1-loop amplitude of the open strings stretched between the two D-particles.
  However, care must be taken to allow only states invariant under
  the orbifold projection
  to propagate in the loop \cite{Sen-1}. 
  The open string states on a non-BPS D-particle are
  labelled by Chan-Paton factors\footnote{This can be derived
  from a D0 anti-D0 system in type IIA following the prescription
  given in \cite{Sen-5,Sen-review}.} $\one$ and $\sigma_1$.
  States with Chan-Paton factor $\one$ have usual GSO projection and are
  even under ${\cal I}_4$. On the other hand, states with Chan-Paton factor
  $\sigma_1$ have opposite GSO projection and are odd under ${\cal I}_4$.
  Moreover,
  the symmetry $(-1)^{F_L}$ does not have any action on the open string
  oscillators.
  Accordingly, the open string 1-loop amplitude is given by:
  \bea
  {\cal A} &=& 2V_1 \int_0^\infty {d \tau \over 2 \tau}
  {\rm Tr}_{\rm NS-R} \left\{ {1\over 4}
  \left( 1 - (-1)^{F+G} \right) \left( 1 + {\cal I}_4 \right)
  e^{-2\pi\tau L_0} \right.
  \nonumber \\
  &&\hspace{3cm}
  + \left. {1 \over 4} \left( 1 + (-1)^{F+G} \right)
  \left( 1 - {\cal I}_4 \right)
  e^{-2\pi\tau L_0} \right\} \nonumber \\
  &=&
  2V_1 \int_0^\infty {d \tau \over 2 \tau}
  {\rm Tr}_{\rm NS-R} \left\{ {1\over 2} \left( 1 - (-1)^{F+G} {\cal I}_4
  \right)
  e^{-2\pi\tau L_0} \right\} \, ,
  \\
  \nonumber
  \eea
  where $F$ and $G$ are the (worldsheet) fermion and superghost number
  operators for the open string.
  The tachyon is projected out in the trace and
  this renders the non-BPS D-particle stable \cite{Bergman-Gaberdiel-1}.
  We obtain the following 1-loop amplitude:
  \bea
  {\cal A} &=&
  {1 \over 2} V_1 (8 \pi^2 \alpha^\prime)^{-1/2}
  \int_0^\infty {d \tau \over \tau^{3/2}} \,
  {\rm e}^{-{ y^2 \tau \over 2\pi \alpha^\prime}}
  \, \times
  \\
  &&\hspace{1cm} \times \left\{
  {f_3^8 ({\rm e}^{-\pi\tau}) - f_2^8 ({\rm e}^{-\pi\tau}) \over
  f_1^8 ({\rm e}^{-\pi\tau})} -
  4 \, {f_4^4 ({\rm e}^{-\pi\tau}) f_3^4 ({\rm e}^{-\pi\tau}) \over
  f_1^4 ({\rm e}^{-\pi\tau}) f_2^4 ({\rm e}^{-\pi\tau})} \right\}
  \, .
  \nonumber \\
  \nonumber
  \eea
  As mentioned above, open-closed string consistency allows us to fix the
  normalisation
  of the boundary state:
  \be
  T_0 = 8 (\alpha^\prime)^{3/2} \pi^{7/2} \, ,\qquad
  Q_0 = 8\pi \sqrt{\pi \alpha^\prime} \, .
  \ee
  The tension of the non-BPS D-particle
  is given by $T_0$
  and $Q_0$ is related to its electric charge.
  Using that in ten dimensions
  $\kappa_{10} = 8 \pi^{7/2} g_s (\alpha^\prime)^2$,
  and that for the orbifold $\kappa_{orb} = \sqrt{2} \kappa_{10}$,
  we find the mass per unit volume of the D-particle:
  \be
  M_0 = {T_0 \over \kappa_{orb}} =
  {1 \over g \sqrt{2 \alpha^\prime}} \, ,
  \ee
  which agrees with the mass of the D-particle found in \cite{Sen-1}.
  In analogy with the BPS D-branes,
  we can define an electric charge with respect to
  the (twisted) R-R field as ${\mu}_0 = \sqrt{2} {Q}_0
  = 8 \pi \sqrt{2\pi \alpha^\prime}$.
  It is interesting to observe that the tension $T_0$ of
  this non-BPS D-particle is the same as that of a BPS D-particle of type
  IIA theory in ten dimensions\footnote{The unstable non-BPS
  D-particle of type IIB has a tension $\sqrt{2}$ times bigger
  than the type IIA BPS D-particle. However, in the orbifold case,
  there is an extra factor $1/2$
  in the open string amplitude coming from the projection operator,
  such that the $\sqrt{2}$ factor is compensated.}.
  However, their charges are different. In fact,
  $\mu_0$ is exactly twice the
  charge of the BPS D-particle of type IIA in six dimensions.

  \subsection{The D-particle in the Compact Orbifold}
  \label{compact-boundary}

  Let us consider now the boundary state of a
  non-BPS D-particle in the case of the compact orbifold,
  i.e. type IIB on $T^4/$\orbifold.
  The coordinates $x^i$, $i=6,7,8,9$, are periodic
  and there are 16 fixed planes instead of one,
  located at $x^i=0, \pi R_i$. We consider a non-BPS D-particle
  located on one of the fixed planes, that we choose
  $x^i=0$, $i=6,7,8,9$, for simplicity.

  The boundary state
  is now constructed upon a groundstate
  with zero winding and zero momentum.
  The only piece modified in the boundary state
  with respect to the non-compact case
  is the bosonic part and the overall normalisation.
  These two modifications will only
  affect the NS-NS part of the boundary state.
  The bosonic part is modified to include Kaluza-Klein modes
  in the directions $i=6,\dots,9$:
  \be
  \ket{D0_X} = \delta^{(5)}(\hat{q}^p - y^p) \,
  \prod_{i=6}^9 \, \sum_{n_i \in \Z}
  {\rm e}^{i {\hat{q}_n}^i {n_i \over R_i}} \,
  {\rm exp}\biggl[-\sum_{n=1}^\infty \frac{1}{n}\,
  \alpha_{-n}\cdot S\cdot
  \tilde{\alpha}_{-n}\biggr]\,
  \ket{k=0,n=0} \, .
  \ee
  The states $\ket{n_i} ={\rm exp} ( i q_n^i {n_i \over R_i})
  \ket{n=0}$ are normalised as
  \be
  \bra{n_i} n^\prime_i \rangle = \Phi_i \, \delta_{n_i \, n^\prime_i} \, ,
  \label{normalisation-kaluza}
  \ee
  where $\Phi_i$ is the self-dual
  volume \cite{Veneziano-etal} which satisfies
  \be
  \lim\limits_{R_i \to \infty} \Phi_i = 2\pi R_i
  \, ,\qquad \lim\limits_{R_i \to 0} \Phi_i = {2\pi \alpha^\prime
  \over R_i} \, .
  \ee
  We consider the new overall normalisation factor as a constant times the
  normalisation for the uncompactified case:
  \be
  \ket{D0,\eta}_{\rm NS,U} = {T_0 \over 2} \, {\cal N} \,
  \ket{D0_X}\, \ket{D0_{\rm gh}}\,{\ket{D0_\psi,\eta}}_{\rm NS}
  \,{\ket{D0_{\rm sgh},\eta}}_{\rm NS} \, ,
  \ee
  The factor ${\cal N}$ can be obtained by open-closed string consistency
  and is such that in the decompactification limit
  $R_i \to \infty$, we recover the non-compact boundary state
  \cite{Frau-etal}.
  The amplitude between two non-BPS D-particles in the closed string channel
  is given by:
  \bea
  \bra{D0} {\cal D} \ket{D0} &=&
  {V_1 \alpha^\prime \pi \over 16} (2\pi^2 \alpha^\prime)^{-5/2}
  \int_0^\infty d \ell \, \ell^{-5/2} \,
  {\rm e}^{-{ y^2 \over 2\pi \alpha^\prime \ell}}
  \, \times
  \label{amplitude-compact-v=0}\\
  &&\hspace{-2cm}\times \left\{
  ({T_0})^2 \, {\cal N}^2
  \left(\prod\limits_{i=6}^9 \Phi_i \right)
  \prod\limits_{i=6}^9 \sum\limits_{n_i \in \Z}
  {\rm e}^{-{\pi \alpha^\prime \ell \over 2} ({n_i \over R_i})^2} \,
  {f_3^8 ({\rm e}^{-\pi\ell}) - f_4^8 ({\rm e}^{-\pi\ell}) \over
  f_1^8 ({\rm e}^{-\pi\ell})} \right.
  \nonumber \\
  &&\hspace{3cm}
  \left.- ({Q_0})^2
  \, {f_2^4 ({\rm e}^{-\pi\ell}) f_3^4 ({\rm e}^{-\pi\ell}) \over
  f_1^4 ({\rm e}^{-\pi\ell}) f_4^4 ({\rm e}^{-\pi\ell})} \right\} \, .
  \nonumber
  \eea
  Using the worldsheet duality $\ell = {1\over \tau}$ and the Poisson
resummation formula
  \be
  \sum\limits_{n \in \Z}
  {\rm e}^{-{\pi \alpha^\prime \ell \over 2} ({n \over R})^2}
  =
  \sqrt{ 2 \over \ell \alpha^\prime} \, R
  \sum\limits_{m \in \Z}
  {\rm e}^{-{2 \pi \over \alpha^\prime \ell } ({m R})^2} \, ,
  \ee
  one obtains
  \bea
  \bra{D0} {\cal D} \ket{D0} &=& {V_1 \alpha^\prime \pi \over 16}
  (2 \pi^2 \alpha^\prime)^{-5/2}
  \int_0^\infty {d \tau \over \tau^{3/2}} \,
  {\rm e}^{-{ y^2 \tau \over 2\pi \alpha^\prime}}
  \, \times
  \\
  && \hspace{-2cm}\times \left\{
  (T_0)^2 {\cal N}^2 \left({2 \over \alpha^\prime}\right)^2
  \left( \prod\limits_{i=6}^9 R_i \Phi_i \right) \,
  \prod\limits_{i=6}^9 \, \sum\limits_{m_i \in \Z}
  {\rm e}^{-{2 \pi \tau \over \alpha^\prime } ({m_i R_i})^2} \,
  {f_3^8 ({\rm e}^{-\pi\tau}) - f_2^8 ({\rm e}^{-\pi\tau}) \over
  f_1^8 ({\rm e}^{-\pi\tau})} \right.
  \nonumber \\
  &&\hspace{3cm}
  \left.- (Q_0)^2 \,
  {f_4^4 ({\rm e}^{-\pi\tau}) f_3^4 ({\rm e}^{-\pi\tau}) \over
  f_1^4 ({\rm e}^{-\pi\tau}) f_2^4 ({\rm e}^{-\pi\tau})} \right\}
  \, .\nonumber
  \\ \nonumber
  \eea
  Open-closed string consistency imposes that this
  amplitude must be equal to the
  open string 1-loop amplitude for the compactified case
  \bea
  {\cal A} &=& 2V_1 \int_0^\infty {d \tau \over 2 \tau}
  {\rm Tr}_{\rm NS-R} \left\{ {1\over 2} \left( 1 + (-1)^F {\cal I}_4 \right)
  {\rm e}^{-2\pi\tau L_0} \right\} \nonumber \\
  &=&
  {1 \over 2} V_1 (8 \pi^2 \alpha^\prime)^{-1/2}
  \int_0^\infty {d \tau \over \tau^{3/2}} \,
  {\rm e}^{-{ y^2 \tau \over 2\pi \alpha^\prime}}
  \, \times
  \\
  &\times& \left\{
  \prod\limits_{i=6}^9 \sum\limits_{m_i \in \Z}
  {\rm e}^{-{2 \pi \tau \over \alpha^\prime } ({m_i R_i})^2} \,
  {f_3^8 ({\rm e}^{-\pi\tau}) - f_2^8 ({\rm e}^{-\pi\tau}) \over
  f_1^8 ({\rm e}^{-\pi\tau})}
  - 4 \,
  {f_4^4 ({\rm e}^{-\pi\tau}) f_3^4 ({\rm e}^{-\pi\tau}) \over
  f_1^4 ({\rm e}^{-\pi\tau}) f_2^4 ({\rm e}^{-\pi\tau})} \right\}
  \, .\nonumber
  \\ \nonumber
  \eea
  This gives the same value for $Q_0$ as in the uncompactified case,
  which is expected
  since the compactification is done in the transverse directions of the
  D-particle.
  This also fixes the normalisation factor ${\cal N}$ to be:
  \be
  {\cal N}= \left(\prod\limits_{i=6}^9 2\pi R_i \Phi_i \right)^{-1/2} \, .
  \ee
  It is easy to check that with this normalisation and using
  $\Phi_i = 2\pi R_i$, in
  the decompactified limit, we can
  recover the untwisted boundary state for the non-compact case.
  Thus the net effect
  of the compactification is a renormalisation of the tension of the
  D-particle and as
  expected, the tension depends on the compactification radii.
  At this point we can
  recover the vanishing of the amplitude at the
  critical radius \cite{Gaberdiel-Sen}.
  The open strings on the D-particle have now winding modes:
  \be
  M^2 = \sum_{i=6}^{9} \left({w_i R_i \over \alpha^\prime}\right)^2 +
  {1 \over \alpha^\prime} \left( N- {1 \over 2} \right) \, .
  \ee
  The groundstate with zero winding is the tachyon and is projected out by the
  orbifold symmetry. At a particular critical value of the radii
  \be
  R_i = \sqrt{ \alpha^\prime \over 2} \, ,\qquad i=6,7,8,9,
  \ee
  there are four states, for which only one $w_i \neq 0$, which become
  massless
  \cite{Sen-5,Majumder-Sen,Bergman-Gaberdiel-2}. These are the modes
  of the tachyon field that at the critical radius correspond to the
  marginal deformation which takes the D-particle to a bound state of a
  D-string
  and an anti-D-string \cite{Sen-1}.
  At this critical radius a Bose-Fermi degeneracy takes
  place, which translates to the fact that the 1-loop amplitude vanishes
  \cite{Gaberdiel-Sen}. In the closed
  string channel this can be seen by using the relations:
  \bea
  \sum\limits_{n \in \Z}
  {\rm e}^{-\pi \ell n^2} &=&
  f_1({\rm e}^{-\pi \ell}) f_3^2 ({\rm e}^{-\pi \ell})
  \\
  f_4({\rm e}^{-\pi \ell}) f_2 ({\rm e}^{-\pi \ell}) f_3 ({\rm e}^{-\pi \ell})
  &=& \sqrt{2} \, . \nonumber
  \eea
  Moreover, the normalisation factor ${\cal N}$ at the critical radius becomes
  \be
  {\cal N} = {1 \over 2 \pi^2 \alpha^\prime}
  \left( \prod\limits_{i=6}^9 \Phi_i \right)^{-1/2} \, .
  \ee
  Hence we can write the amplitude as
  \bea
  \bra{D0} {\cal D} \ket{D0} &=&
  {V_1 \alpha^\prime \pi \over 16} (2\pi^2 \alpha^\prime)^{-5/2}
  \int_0^\infty d \ell \, \ell^{-5/2} \,
  {\rm e}^{-{ y^2 \over 2\pi \alpha^\prime \ell}}
  \, \times
  \nonumber \\
  &&\hspace{-2cm}\times
  {f_3^4 ({\rm e}^{-\pi\ell}) \over
  f_1^4 ({\rm e}^{-\pi\ell}) f_2^4 ({\rm e}^{-\pi\ell})
  f_4^4 ({\rm e}^{-\pi\ell})}
  \left\{
  \left({T_0 \over \pi^2 \alpha^\prime}\right)^2 \,
  \left( {f_3^8 ({\rm e}^{-\pi\ell}) - f_4^8 ({\rm e}^{-\pi\ell})} \right)
  \right.
  \nonumber \\
  &&\hspace{3cm}
  \left.- ({Q_0})^2 \,
  f_2^8 ({\rm e}^{-\pi\ell}) \right\} \, .
  \label{bose-fermi-deg}
  \\ \nonumber
  \eea
  which vanishes, at any distance $y$, by the abstruse identity and using the
  expressions for $T_0$ and $Q_0$. In the open string channel it is simpler to
  demonstrate this by using the above identities.

\subsection{Long and Short Distance Interactions}

In this Section we  consider the interaction amplitude for non-BPS D-particles
in relative motion.
This will give a velocity dependent potential which we will compare with the
usual BPS case. This amplitude can be obtained by using a boosted
boundary state \cite{Divecchia-4} for the non-BPS D-particle.
For definiteness we consider a D-particle moving in the direction
$X^1$ with a velocity $v$, interacting with another non-BPS D-particle
at rest. We consider first the non-compact orbifold. 
The boosted boundary state include some 
$v$-dependent modifications that we write explicitly below.
In the NS-NS untwisted sector, the bosonic part of the boundary state becomes
\be
\ket{D0_X,v} = \sqrt{1-v^2} 
\left( \prod\limits_{i=2}^9 \delta({\hat q}^p) \right)
\, \delta({\hat q}^1 + v {\hat q}^0) \, 
{\rm exp}\biggl[-\sum_{n=1}^\infty \frac{1}{n}\,
  \alpha_{-n}\cdot S(v) \cdot
  \tilde{\alpha}_{-n}\biggr]\,
  \ket{k=0} \, ,
\ee
where the matrix $S(v)$ is given by
\be
S_{00}(v)=S_{11}(v)= - {1 + v^2 \over 1 - v^2} \, ,\qquad
S_{10}(v)=S_{01}(v)= - { 2v \over 1 - v^2} \, ,
\label{matrix-S(v)}
\ee
and $S(v) = S$ for the other components.
The other pieces in the NS-NS untwisted  boundary state have the usual form
except for a substitution of the matrix $S$ by $S(v)$.
Similarly in the R-R twisted sector, the boosted bosonic boundary state
takes the form:
\begin{displaymath}
\ket{D0_X,v}_{\rm T} = \sqrt{1-v^2} \prod\limits_{p=2}^5 
\delta({\hat q}^p)
\, \delta({\hat q}^1 + v {\hat q}^0) \, 
{\rm exp}\biggl[-\sum_{t.m.} \frac{1}{n}\,
  \alpha_{-n}\cdot S(v) \cdot
  \tilde{\alpha}_{-n}\biggr]\,
  \ket{k=0,n=0} \, ,
\end{displaymath}
where with $t.m.$ we indicate that one takes the corresponding
{\it twisted moddings} of the twisted sector.
For the other pieces we do the same substitution $S \rightarrow S(v)$.
Finally, for the R-R zero-mode part there is as well a 
$v$-dependent modification:
\be
\ket{D0,\eta,v}_{\rm R,T}^{(0)} = M_{ab}(v) \ket{a}_{\rm T} 
\widetilde{\ket{b}}_{\rm T} \, ,
\ee
and the matrix $M_{ab}(v)$ is given in Appendix A.
We define the distance between the particles as 
$r^2 = b^2 + t^2 v^2 (1-v^2)^{-1}$, where $t$ is the proper time of the moving
particle along which we also integrate; $b$ is the impact parameter:
$b^2 = y_2^2 + \dots y_5^2$. 
For convenience we also define the following variable
\be
u = {1 \over 2\pi i} \, {\rm ln} \, {1-v \over 1+v} \, .
\ee
Finally, the cylinder amplitude takes the following form:
\bea
\bra{D0,v} {\cal D} \ket{D0} &=&
(8 \pi^2 \alpha^\prime)^{-1/2} \, {\rm sin} \, \pi u
\, \int\limits_0^\infty d\ell \, \ell^{-{9 \over 2}} \, 
\int\limits_{-\infty}^\infty d t \, {\rm e}^{- 
{r^2 \over 2\pi\alpha^\prime \ell}} \, \times \, 
\label{amplitude-non-compact-v}
\\
&&\hspace{-3cm} \left\{ 
{\Theta_3 (u,i\ell) f_3^6({\rm e}^{-\pi\ell}) 
- \Theta_4 (u,i\ell) f_4^6({\rm e}^{-\pi\ell})
\over \Theta_1 (u,i\ell) f_1^6({\rm e}^{-\pi\ell})}
\, - \, 4 \ell^2 \, 
{ \Theta_2 (u,i\ell) f_2^2({\rm e}^{-\pi\ell})f_3^4({\rm e}^{-\pi\ell})
\over \Theta_1 (u,i\ell) f_1^2({\rm e}^{-\pi\ell})
f_4^4({\rm e}^{-\pi\ell})} \right\} \, ,\nonumber \\
\nonumber
\eea
where $\Theta_s (u,i\ell)$ are the Jacobi $\Theta$-functions,
and we have used the value of the tension and charge of the D-particle.
For  $v=0$ we recover the amplitude for the static interaction given in 
(\ref{amplitude-non-compact-v=0}).

In order to study the interaction we define the
interacting potential of the scattering $\cal{U}$, in the following way:
\be
\bra{D0,v} {\cal D} \ket{D0} = \int\limits_{-\infty}^\infty  dt
\,\,  {\cal U} (v,r(t)) \, .
\ee
We can extract the long range interaction potential
taking the limit $\ell \to \infty$ in the integrand
of (\ref{amplitude-non-compact-v}), and then performing the
integral in $\ell$. For slow velocities, we find the following 
expansion in powers of $v$: 
\be
{\cal U}_{closed} \simeq
{ (2 \pi \alpha^\prime)^3 \over (4\pi)^{1/2}}
\, 8 \, \left\{ (1 + {1\over 2} v^2 + {1 \over 24} v^4) \right.
\left( {\Gamma({7\over 2}) \over r^7} - 
{1 \over (2\pi\alpha^\prime)^2}\, {\Gamma({3\over 2}) \over r^3} \right)
+ \left. {1 \over 8} \, v^4 \, {\Gamma({7\over 2}) \over r^7} \right\}
\, ,
\label{potential-non-compact-closed}
\ee
to order $v^4$.
There is a static force, as expected since the D-particle is non-BPS. 
Moreover, the 
velocity dependent corrections start at order $v^2$,
as for the potential between two BPS D-branes of different dimensionality
\cite{Lifschytz,Tseytlin}, and as for the non-BPS D-particle of type I
\cite{Gallot-etal}.
This is something we could expect from the fact that the particle
breaks all supersymmetries.

We can analyse this amplitude for very short distances. In this region
the open string description dominates. Using the modular properties of the
$\Theta$-functions we can write
(\ref{amplitude-non-compact-v}) in the open string channel:
\bea
\bra{D0,v} {\cal D} \ket{D0} &=& - \, i 
(8 \pi^2 \alpha^\prime)^{-1/2} \, {\rm sin} \, \pi u
\, \int\limits_0^\infty d \tau \, \tau^{-{1 \over 2}} \, 
\int\limits_{-\infty}^\infty d t \, {\rm e}^{- 
{r^2 \over 2\pi\alpha^\prime} \tau} \, \times \, 
\label{open-amplitude-non-compact-v}
\\
&&\hspace{-3cm} \left\{ 
{\Theta_3 (-iu\tau,i\tau) f_3^6({\rm e}^{-\pi\tau}) 
- \Theta_2 (-iu\tau,i\tau) f_2^6({\rm e}^{-\pi\tau})
\over \Theta_1 (-iu\tau,i\tau) f_1^6({\rm e}^{-\pi\tau})}
\, - \, 4 \,  
{ \Theta_4 (-iu\tau,i\tau) f_4^2({\rm e}^{-\pi\tau})f_3^4({\rm e}^{-\pi\tau})
\over \Theta_1 (-iu\tau,i\tau) f_1^2({\rm e}^{-\pi\tau})
f_2^4({\rm e}^{-\pi\tau})} \right\} \, .\nonumber \\
\nonumber
\eea
We can derive the interaction potential
at short distances by taking the limit $\tau \to \infty$ in the integrand
and then performing the
integral in $\tau$. Note that
the pieces originally from the NS-NS and R-R sectors 
multiply now in (\ref{open-amplitude-non-compact-v})
the same powers of $\tau$, hence we can expect 
that the interaction potential will be very different from the closed
string case. This time we obtain:
\be
{\cal U}_{open} = 
- \, i (8\pi^2 \alpha^\prime)^{-1/2} {\rm sin} \, \pi u
\int\limits_0^{\infty} d\tau \, \tau^{-1/2} 
{\rm e}^{- {r^2 \over 2\pi\alpha^\prime} \tau} \, 
{ 2 + 2 {\rm cos} (-2\pi i u \tau) - 8 {\rm cos} ( -i\pi u \tau)
\over {\rm sin} (-i \pi u \tau) } \, ,
\ee
which for slow velocities and after integrating in $\tau$ becomes
(to order $v^4$):
\be
{\cal U}_{open} = {4 \over 2\pi \alpha^\prime} \, r \, + \,
{(2\pi \alpha^\prime)^3 \, \Gamma({7\over 2}) \over (4\pi)^{1/2}}
\, {v^4 \over r^7} \, .
\label{potential-non-compact-open}
\ee
Surprisingly, there is no $v^2$ correction to the potential,
as for BPS D-branes.
The first term is the linear repulsive force coming from the stretched 
strings. Moreover, the $v^4$ correction
has precisely  the same form as the $v^4$ term for a scattering
of two BPS D-particles. For BPS branes the $v^4$ term in the open string
channel coincides with the $v^4$ term from the closed string description
\cite{DKPS}. This does not occur in the non-compact orbifold. As we will
see below for the compact case,
even though
at the critical radius the static potential and
$v^2$ terms will be  suppressed, 
the matching of the $v^4$ terms will not occur either. 
Hence the BPS-like behaviour will not extend beyond order $v^2$.

We extend next the analysis of the cylinder amplitude for the case 
of relative motion of the D-particles in the compact orbifold.
We consider again one non-BPS D-particle moving in the direction
$X^1$ with a velocity $v$, interacting with another non-BPS D-particle
at rest. 
The boosted boundary state include the
$v$-dependent modifications on top of the modifications
due to compactification.
In the NS-NS untwisted sector, the bosonic part of the boundary state becomes
\bea
\ket{D0_X,v} &=& \sqrt{1-v^2} 
\left( \prod\limits_{p=2}^5 \delta({\hat q}^p) \right)
\, \delta({\hat q}^1 + v {\hat q}^0) \, \times
\\
&&\hspace{-2cm}\times \, 
\left( \prod\limits_{i=6}^9 \sum\limits_{n_i \in \Z} 
{\rm e}^{i {\hat q}^i_n {n^i \over R_i}} \right) \,
{\rm exp}\biggl[-\sum_{n=1}^\infty \frac{1}{n}\,
  \alpha_{-n}\cdot S(v) \cdot
  \tilde{\alpha}_{-n}\biggr]\,
  \ket{k=0,n=0} \, ,
\eea
where the matrix $S(v)$ is as given before in (\ref{matrix-S(v)}).
In the R-R twisted sector, the boosted bosonic boundary state
and fermionic zero-mode have the same form as for the non-compact case,
and the other pieces have the usual form
except for a substitution of the matrix $S$ by $S(v)$.
Using the same notation as before, 
the cylinder amplitude takes the following form:
\bea
\bra{D0,v} {\cal D} \ket{D0} &=&
\sqrt{2 \over \pi^2 \alpha^\prime} \, {\rm sin} \, \pi u
\, \int\limits_0^\infty d\ell \, \ell^{-{5 \over 2}} \, 
\int\limits_{-\infty}^\infty d t \, {\rm e}^{- 
{r^2 \over 2\pi\alpha^\prime \ell}} \, \times \, 
\label{amplitude-compact-v}
\\
&&\hspace{-2.5cm} \left\{ \left({\alpha^\prime \over 4} \right)^2 
\prod\limits_{i=6}^9 R_i^{-1}
\left( \prod\limits_{i=6}^9 \sum\limits_{n_i \in \Z} 
{\rm e}^{- {\pi \alpha^\prime\ell \over 2} ({n_i \over R_i})^2} \right)
\left( {\Theta_3 (u,i\ell) f_3^6({\rm e}^{-\pi\ell}) 
- \Theta_4 (u,i\ell) f_4^6({\rm e}^{-\pi\ell})
\over \Theta_1 (u,i\ell) f_1^6({\rm e}^{-\pi\ell})} \right) \right.
\nonumber \\
&&\hspace{3cm}- \, 
\left. { \Theta_2 (u,i\ell) f_2^2({\rm e}^{-\pi\ell})f_3^4({\rm e}^{-\pi\ell})
\over \Theta_1 (u,i\ell) f_1^2({\rm e}^{-\pi\ell})
f_4^4({\rm e}^{-\pi\ell})} \right\} \, .\nonumber \\
\nonumber
\eea
For $v=0$ we recover the amplitude for the static interaction given in 
(\ref{amplitude-compact-v=0}).
For slow velocities ($u \simeq (i\pi)^{-1} v$) and
after integration in $\ell$, the long range 
interaction potential to order $v^4$ is given by
\be
{\cal U}_{closed} \simeq {4\pi \alpha^\prime \over r^3}
\left\{ (1 + {1 \over 2} v^2 + {1 \over 6} v^4) 
\left( {\alpha^{\prime 2} \over 4} \prod_{i=6}^9 R_i^{-1}
-1 \right) -  {v^4 \over 8} \right\} \, .
\label{scattering-potential}
\ee
We see that for generic radii the potential has $v^2$ corrections,
as before, which is typical for potentials between 
two BPS D-branes of different dimensionality
\cite{Lifschytz,Tseytlin}, and occurs also for the non-BPS D-particle of type I
\cite{Gallot-etal}. On the other hand,
at the critical radius they start at $v^4$, as for BPS D-branes
\cite{Lifschytz,Bachas}.
Notice that the static and $v^2$ terms of the potential also vanishes
for other values of the radii. However, those radii do not make
the amplitude (\ref{amplitude-compact-v=0}) vanish. Moreover, this
would require some of the radii to be below the critical radius, and
the D-particle would not be stable.

In order to study the short-distance behaviour we
write the scattering amplitude in the open channel:
\bea
\bra{D0,v} {\cal D} \ket{D0} &=&
- \, i \sqrt{2 \over \pi^2 \alpha^\prime} \, {\rm sin} \pi u
\, \int\limits_0^\infty d\tau \, \tau^{-{1 \over 2}} \, 
\int\limits_{-\infty}^\infty d t \, {\rm e}^{- 
{r^2 \over 2\pi\alpha^\prime}\tau} \, \times \, 
\\
&&\hspace{-3cm} \left\{ {1 \over 4} 
\left( \prod\limits_{i=6}^9 \sum\limits_{m_i \in \Z} 
{\rm e}^{- {2 \pi \tau \over \alpha^\prime} ({m_i R_i})^2} \right)
\left( {\Theta_3 (-iu\tau,i\tau) f_3^6({\rm e}^{-\pi\tau}) 
- \Theta_2 (-iu\tau,i\tau) f_2^6({\rm e}^{-\pi\tau})
\over \Theta_1 (-iu\tau,i\tau) f_1^6({\rm e}^{-\pi\tau})} \right) \right.
\nonumber \\
&&\hspace{3.5cm}- \, 
\left. { \Theta_4 (-iu\tau,i\tau) f_4^2({\rm e}^{-\pi\tau})
f_3^4({\rm e}^{-\pi\tau})
\over \Theta_1 (-iu\tau,i\tau) f_1^2({\rm e}^{-\pi\tau})
f_2^4({\rm e}^{-\pi\tau})} \right\} \, .\nonumber \\
\nonumber
\eea
To extract the leading contribution at short distances and low velocities we
take the limit $\tau \to \infty$ in the integrand. This time
one must also include contributions with one unit of winding number 
$m_i=1$, which are relevant very close to the critical radius
$R_i \geq R_c$:
\bea
{\cal U}_{open} &=&
\sqrt{2 \over \pi^2 \alpha^\prime}
\int\limits_0^\infty d\tau \, \tau^{-1/2}
\, {\rm e}^{- {r^2 \over 2\pi\alpha^\prime}\tau} \,
{{\rm sin} \pi u \over 2i {\rm sin} ( -i\pi u \tau)} \,
\times
\\
&&\times \, 
\left( {\rm cos}( 2 \pi i u \tau) - 4 {\rm cos} (i \pi u \tau)
+ 1 + {1 \over 2} \sum\limits_{i=6}^9 
{\rm e}^{- {2\pi R_i^2 \over \alpha^\prime} \tau + \pi \tau} \right) \, .
\nonumber
\eea
At low velocities the expression simplify. 
Carrying out the integral in $\tau$ we find:
\be
{\cal U}_{open} =
{4 \over 2\pi\alpha^\prime}
\left( \, r - {1 \over 4}\sum\limits_{i=6}^9
\sqrt{ r^2 + 4\pi^2 (R_i^2 - {\alpha^\prime \over 2}) } \, \right)
+ {(2\pi\alpha^\prime)^3 \, \Gamma({7 \over 2}) \over (4\pi)^{1/2}} 
\, {v^4 \over r^7} \, .
\label{scattering-potential-open}
\ee
As for the non-compact case we find no $v^2$ corrections, and the $v^4$
corrections are again the same as for the BPS D-particle in ten dimensions.
Notice also that the $r$-dependence of the $v^4$ term is as for the 
uncompactified case.
The static part of the potential vanishes when all the radii are equal to 
the critical radius.
Remarkably, at the critical radius ${\cal U}_{open}$ takes
a very BPS-like form. However, as we announced before,
the $v^4$ terms does not match with the closed string result.

From this analysis we conclude that the long and short range 
interactions of the non-BPS D-particle are quite different for generic radii
of the orbifold. In the particular case of the critical radius the static
and $v^2$ terms of the interaction potential are absent in the open and 
closed string description. At the critical radius,
to order $v^2$, we expect an equivalence between the 
(super)gravity and worldvolume description of the non-BPS D-particle.
At order $v^4$ the open strings begin to describe the 
geometry of non-BPS D-particle very differently from the closed strings.


  \section{Spacetime Description}


  As shown in \cite{Divecchia-1},
  the boundary state of a D-brane encodes the long distance behaviour of the
  corresponding classical solution of supergravity.
  At large distances, this classical solution tends to a
  flat background configuration. The fluctuations around this background
  is exactly reproduced by the boundary state.
  From the NS-NS part one obtains the asymptotic
  behaviour of the dilaton and the metric, and from the R-R part one
  obtains the asymptotic behaviour
  of the R-R form potential under which the brane is charged.
  In this section we implement this technique
  on the stable non-BPS D-particle to obtain 
  the asymptotic form of the solution.
  We obtain a metric and a dilaton in the bulk, which depend on all the
  coordinates transverse to the D-particle, and whose dependence is the
  expected one for a particle in 10 dimensions. On the other hand, we also
  find a twisted R-R 1-form which is restricted to the fixed plane and
  whose dependence on the spatial coordinates on the fixed plane is
  the correct one for a particle in 6 dimensions.
  We describe first the non-compact case.
	
  \subsection{The Non-compact Case}

  Given a certain massless closed string state $\ket{\varphi}$, normalised
  as $\bra{\varphi} {\varphi} \rangle = \varphi^2$, one can define
  a projection operator $\bra{P_{(\varphi)}}$ associated to this state, such
  that  $\bra{P_{(\varphi)}} {\varphi} \rangle = \varphi$
  \cite{Divecchia-3}.
  The asymptotic behaviour of the classical field
  $\varphi$, generated
  by a D-brane is determined by computing
  the overlap of the boundary state with the state $\ket{P_{(\varphi)}}$
  with the insertion of a closed string
  propagator \cite{Divecchia-1,Divecchia-3}:
  \be
  \bra{P_{(\varphi)}} {\cal D} \ket{D} \, .
  \label{fluctuation-amplitude}
  \ee
  Since the D-particle has an untwisted NS-NS part and a twisted R-R part,
  the D-particle will excite gravitons $g_{\mu\nu}$ and dilatons $\phi$
  in the untwisted sector
  and a R-R 1-form field ${\cal C}^{(1)}$ in the twisted sector.
  At large distances, these fields will be of the form of a
  fluctuation around a flat
  background\footnote{We use the same normalisation as in
  \cite{Divecchia-1}.}:
  \be
  g_{\mu\nu} \simeq \eta_{\mu\nu} + 2 \kappa_{10} \, \delta h_{\mu\nu} \,
  ,\qquad
  \phi \simeq \kappa_{10} \sqrt{2} \, \delta \phi \, ,\qquad
  {\cal C}^{(1)} \simeq \kappa_{6} \sqrt{2} \, \delta C^{(1)} \, .
  \label{generic-fluctuation}
  \ee
  The fluctuations $\delta h_{\mu\nu}$, $\delta \phi$ and $\delta C^{(1)}$,
  are given by (\ref{fluctuation-amplitude})
  using the state $\ket{D0}$ constructed previously.
  We consider first the case of the uncompactified orbifold.
  In the untwisted sector, only those which are
  invariant under the orbifold symmetry
  will appear in the theory. This implies that string states which depend
  on the momentum in the orbifolded directions are only invariant if they
  are symmetric with respect to ${\cal I}_4:$  $k^i \rightarrow -k^i$,
  $i=6,7,8,9$. In particular, these will include
  gravitons that propagate off the orbifold fixed plane which are 
symmetric in $k^i$.

  The projection operators in the
  untwisted NS-NS sector are given in the $(-1,-1)$ picture.
  For simplicity, we use the following notation for the
  NS-NS groundstate in the $(-1,-1)$ picture:
  \be
  \ket{k} \equiv \ket{k/2}_{-1} \widetilde{\ket{k/2}}_{-1} \, .
  \ee
  The projectors for the dilaton and the metric
  are given by:
  \bea
  \bra{P_{(\phi)}} &=&
  {1\over 4\sqrt{2}} \left( \bra{{k_\bot},k^i} +
  \bra{{k_\bot},- k^i} \right)
  \, \psi^\nu_{1/2} \tilde\psi^\mu_{1/2}
  \,
  \left( \eta_{\mu\nu} - k_\mu \ell_\nu - k_\nu \ell_\mu \right)
  \, ,\nonumber \\
  \bra{P_{(h)}^{\mu\nu}} &=&
  {1\over 4} \left( \bra{{k_\bot},k^i} +
  \bra{{k_\bot},- k^i} \right)
  \, \left( \psi^\nu_{1/2} \tilde\psi^\mu_{1/2} +
  \psi^\mu_{1/2} \tilde\psi^\nu_{1/2} \right)
\\
&&  - \bra{P_{(\phi)}} {1 \over 2\sqrt{2}} 
\left( \eta^{\mu\nu} -k^\mu \ell^\nu - k^\nu \ell^\mu \right) \, ,\nonumber
  \eea
  where $k=(k_\bot,k^i)$,
  with $k_{\bot}$ the spatial components of the momentum
  along the fixed plane directions.
  Moreover, $\ell$ is such that $\ell^2 =0$ and $k \cdot \ell= 1$.

  In the R-R sector, the most natural picture is the
  asymmetric\footnote{See \cite{Divecchia-1} for an explanation about this.} 
picture $(-1/2,-3/2)$, where the R-R vertex
operator couples to the potential instead of the field strength
\cite{Bianchi-etal,Divecchia-2}.
  In analogy to the type II case described in \cite{Divecchia-2} 
and using $a, b$ as the SO(1,5) spinor
indices, we write the following
  quantum state in the $(-1/2,-3/2)$ picture associated to the twisted R-R
  1-form field\footnote{For simplicity we omit the subindex T of the twisted
  groundstate.}
  \begin{displaymath}
  \ket{C^{(1)}} = {1 \over \sqrt{2}} C^{(1)}_\alpha
  \left\{ \left( {\cal C} \gamma^\alpha \Pi_+\right)_{ab}
  {\rm cos} \gamma_0 \tilde\beta_0
  + \left({\cal C} \gamma^\alpha \Pi_-\right)_{ab}
  {\rm sin} \gamma_0 \tilde\beta_0
  \right\} {\ket{a, {k/2}}}_{-1/2}
  \widetilde{\ket{b, {k/2}}}_{-3/2} \, ,
  \end{displaymath}
  whose conjugate state is given by:
  \begin{displaymath}
  \bra{C^{(1)}} = - C^{(1)}_\alpha \, {1 \over \sqrt{2}} \,
  {}_{-1/2}\widetilde{\bra{b, {k/2}}}
  {}_{-3/2}\bra{a, {k/2}}
  \, \lbrace \left( {\cal C} \gamma^\alpha \Pi_- \right)_{ab}
  {\rm cos} \beta_0 \tilde\gamma_0
  + \left({\cal C} \gamma^\alpha \Pi_+\right)_{ab} {\rm sin}
  \beta_0 \tilde\gamma_0
  \rbrace \, ,
  \end{displaymath}
  with
  \be
  \Pi_\pm = {1 \over 2}
  \left( \one_8 \pm \gamma \right) \, ,
  \ee
  the chirality projector in 6 dimensions, where the gamma-matrix
  $\gamma$ is defined in Appendix \ref{appendixA}.
  This state is normalised
  as $\bra{C^{(1)}} {C^{(1)}}\rangle = C^{(1)}_\alpha C^{(1)\,\alpha}$.
  The corresponding projector is given by
  \begin{displaymath}
  \bra{P_{(C)}{}^\alpha} = - {1 \over \sqrt{2}}
  \, {}_{-1/2}\widetilde{\bra{b, {k_\bot/2}}}
  {}_{-3/2}\bra{a, {k_\bot/2}}
  \lbrace \left( {\cal C} \gamma^\alpha \Pi_- \right)_{ab}
  {\rm cos} \beta_0 \tilde\gamma_0
  + \left({\cal C} \gamma^\alpha \Pi_+\right)_{ab} {\rm sin}
  \beta_0 \tilde\gamma_0
  \rbrace \, .
  \end{displaymath}
  such that $\bra{P_{(C)}{}_\alpha} C^{(1)} \rangle = C^{(1)}_\alpha$.
  The form of this projector is similar to the case of the type IIA
  BPS D-particle,
  with the difference that the twisted R-R groundstate carries spinor indices
  of $SO(1,5)$.

  The NS-NS (R-R) fields have non-zero overlap with the NS-NS
  (R-R) boundary state only.
  We find the following results for the asymptotic fields in momentum space:
  \bea
  \delta \phi (k) &=& \bra{P_{(\phi)}}
  {\cal D}_{a=1/2} \ket{D0}_{\rm NS,U} =
  {T}_0 \, {3 \over \sqrt{8}} \, {V_1 \over k^2} \, ,
  \\
  \delta h_{\mu\nu} (k) &=& \bra{P_{(h)\, \mu\nu}} {\cal D}_{a=1/2}
  \ket{D0}_{\rm NS,U} = {{T}_0 \over 2} \, {V_1 \over k^2} \,
  {\rm diag} \left( {7\over 4}, {1\over 4}, \dots , {1\over 4} \right) \, .
  \eea
  For the twisted R-R field we only have a contribution along the worldline
  direction of the D-particle, namely:
  \be
  \delta C^{(1)}_0 (k_\bot) 
= \bra{P_{(C) \, 0}} {\cal D}_{a=0} \ket{D0}_{\rm R,T} =
  - {Q_0 \over \sqrt{2}} \, {V_1 \over k_{\bot}^2} \, .
  \ee
  Note that
  $\delta C^{(1)}_0$ depends only on the momenta transverse to the D-particle,
  longitudinal to the fixed plane.
  We now make use of the following Fourier transformation in order to
  translate
  the results into position space. For a generic momentum ${K}$, with
  $d-1$ non-zero components, we have:
  \be
  \int dt \, d^{(d-1)} x \,
  {e^{ {\rm i} {K} \cdot {x}} \over (d-3) \,
  \Omega_{d-2} \, |{x}|^{(d-3)}} \,
  = \, {V_1 \over {K}^2} \, ,
  \label{Fourier-1}
  \ee
  where $x$ are $d-1$ spatial coordinates transverse to the D-particle
  and $\Omega_{d-2}$ is the area of a unit sphere surrounding the D-particle.
  For the NS-NS sector we have $d=9$, hence we find the
  following asymptotic behaviour for the
  metric and dilaton:
  \bea
  \delta \phi (x) &=& {3 T_0 \over 14\sqrt{2} \, \Omega_8}
  \, {1 \over |x|^7 } \, ,
  \nonumber \\
  \delta h_{\mu\nu} (x) &=& {T_0 \over 14 \, \Omega_8}
  \, {1 \over |x|^7 } \, {\rm diag}
  \left( {7\over 4}, {1 \over 4} , \dots , {1 \over 4} \right) \, .
  \label{NSNS-asymptotics}
  \eea
  In the R-R sector, since the momentum dependence is restricted to
  the fixed plane spatial directions, we have $d=6$,
  and the asymptotic form of the R-R 1-form is found to be:
  \be
  \delta C^{(1)}(y) = - {{Q}_0 \over 3\sqrt{2} \, \Omega_4} \,
  {1 \over |y|^3} \, ,
  \label{RR-asymptotics}
  \ee
  where now $y$ indicates the spatial directions in the fixed plane only,
  and $\Omega_4$ is the volume of a unit 4-sphere surrounding the particle
  inside the fixed plane.
  Note that the power of $|y|$ is exactly the power expected for
  a particle in 6 dimensions.
  We see, however, that the untwisted fields do
  {\it see} the entire spacetime.
  Thus, although the D-particle is stuck on the fixed plane,
  the associated metric and dilaton background solution
  in the uncompactified case
  will extend also in the orbifolded directions. This is a consequence of the
  fact that the D-particle can emit massless untwisted fields off the fixed
  plane
  which are symmetrised in the momenta along the orbifolded directions.

  \subsection{The Compact Case}

  In order to consider a spacetime description for a stable non-BPS D-brane,
  in a regime where it remains valid classically,
  one should make sense of a superposition of them,
  which turns out to be possible only in the compact orbifold
  at the critical radius. In this section we extend
  the analysis of the metric and dilaton of the D-particle to
  the case in which the space transverse
  to the orbifold fixed plane is a 4-torus,
  i.e. type IIB on $T^4/$ \orbifold.

  We derive first the asymptotic form of the metric and dilaton
  in the compactified case using an approximation in the background fields,
  that we will compare with the boundary state calculation.
  Consider one of the compact orbifolded directions:
  $x_9 \sim x_9 + 2\pi R_9$.
  A D-particle sitting at the origin of this $S^1$ can be seen
  from the point of view of the covering space as an infinite
  array of equally spaced
  D-particles. Thus we can write:
  \be
  {1 \over |x|^7} \simeq \sum\limits_{n \in \Z}
  {1 \over \left(r^2 +(x_9-2\pi n R_9)^2 \right)^{7/2}}
  \, ,
  \ee
  with $|x|^2=r^2 + x_9^2$, and $r^2=x_1^2+ \dots + x_8^2$.
  We can approximate the sum by an integral
  assuming that the distance in the non-compact
  directions is much larger that the size of the compact one,
  i.e. $r \gg R_9$. Changing variables we can write
  \be
  \sum\limits_{n \in \Z}
  {1 \over \left(r^2 +(x_9-2\pi n R_9)^2 \right)^{7/2}}
  \simeq {1 \over 2\pi R_9 r^6} \int\limits_{-\infty}^{+\infty}
  du {1 \over (1+u^2)^{7/2}} = {I_5 \over 2\pi R_9} {1 \over r^6} \, ,
  \ee
  where we use the notation:
  \be
  I_n = \int_0^\pi d\theta \, {\rm sin}^n \theta \, .
  \ee
  Moreover, these integrals satisfy the properties
  $n I_n = (n-1) I_{n-2}$, for $n \geq 2$, and
  \be
  \Omega_d = 2 I_{d-1} I_{d-2} \cdots I_1 I_0 \, .
  \ee
  If we do this approximation for each of the compactified directions
  of the orbifold we obtain
  \be
  {1 \over |x|^7} \simeq
  I_2 I_3 I_4 I_5 \, \prod\limits_{i=6}^9 (2\pi R_i)^{-1}
  \, {1 \over |y|^3} \, ,
  \label{approximation-2}
  \ee
  with $y$ now indicating the spatial directions in the fixed plane.
  Within this approximation we obtain the following asymptotic behaviour for
  the
  dilaton and metric for the compactified case:
  \bea
  \delta \phi (y) &\simeq& {T_0 \over 2\sqrt{2} \, \Omega_4}
  \, \prod_{i=6}^9 (2\pi R_i)^{-1} \, {1 \over |y|^3} \, ,
  \nonumber \\
  \delta h_{\mu\nu} (y) &\simeq& {T_0 \over 6 \, \Omega_4}
  \, \prod_{i=6}^9 (2\pi R_i)^{-1} \, {1 \over |y|^3} \, {\rm diag}
  \left( {7\over 4}, {1 \over 4} , \dots , {1 \over 4} \right) \, .
  \label{NSNS-asymptotics-2}
  \eea
  We rewrite the fluctuations using the 6-dimensional
  gravitational constant
  \bea
  g_{\mu\nu} \simeq \eta_{\mu\nu} + 2 \kappa_{10} \, \delta h_{\mu\nu}
  &\equiv& \eta_{\mu\nu} + 2 \kappa_{6} \, {\delta \bar{h}}_{\mu\nu}
  \, ,\nonumber \\
  \phi \simeq \kappa_{10} \sqrt{2} \, \delta \phi
  &\equiv& \kappa_{6} \sqrt{2} \, \delta \bar{\phi}\, , \\
  \delta {C}^{(1)}
  &\equiv& \delta \bar{C}^{(1)} \, .\nonumber
  \label{generic-fluctuation-6-dim}
  \eea
  Using the relation\footnote{We use that for any dimension
  $D$, $2\kappa_D^2 = 16\pi G_D$, and for any lower dimension
  $D-d$, $G_{D-d} = G_D/V_d$, with $V_d$ the volume of the compact space.}
  between $\kappa_{10}$ and $\kappa_{6}$:
  $\kappa_{10} = \kappa_6 \prod\limits_{i=6}^9 (2\pi R_i)^{1/2}$,
  we obtain
  \bea
  \delta \bar{\phi} (y) &\simeq& {T_0 \over 2\sqrt{2} \, \Omega_4}
  \, \prod_{i=6}^9 (2\pi R_i)^{-1/2} \, {1 \over |y|^3} \, ,
  \nonumber \\
  \delta \bar{h}_{\mu\nu} (y) &\simeq& {T_0 \over 6 \, \Omega_4}
  \, \prod_{i=6}^9 (2\pi R_i)^{-1/2} \, {1 \over |y|^3} \, {\rm diag}
  \left( {7\over 4}, {1 \over 4} , \dots , {1 \over 4} \right) \, .
  \label{NSNS-asymptotics-3}
  \eea
  The asymptotic behaviour of the twisted R-R 1-form stays the same as for the
  non-compact case (\ref{RR-asymptotics}).

  We compare now this approximation with the
  asymptotic behaviour derived from the boundary state.
  The closed string fields have an infinite number of Kaluza-Klein
  modes, since the momentum is quantised in the orbifolded directions.
  The massless closed string states excited by the D-particle
  correspond to the massless Kaluza-Klein modes.
  Using the same notation as before and using bars for the objects
  in the compact case, we can write the
  projection operators for the massless states as follows:
  \bea
  {\bra{\bar{P}_{(\phi)}}} &=&
  {1 \over 2\sqrt{2}}
  \prod\limits_{i=6}^9 \Phi_i^{-1/2}
  {\bra{{k_\bot},n_i=0}} \, \psi^\nu_{1/2} \tilde\psi^\mu_{1/2}
  \, \left( \eta_{\mu\nu} -
  (\delta_\mu{}^\alpha \delta_\nu{}^\beta +
  \delta_\mu{}^\beta\delta_\nu{}^\alpha)
  k_\alpha \ell_\beta \right)
  \, ,\nonumber \\
  {\bra{\bar{P}_{(h)}^{\mu\nu}}} &=&
  {1\over 2} \prod\limits_{i=6}^9 \Phi_i^{-1/2} \bra{{k_\bot},n^i=0}
  \, \left( \psi^\nu_{1/2} \tilde\psi^\mu_{1/2} +
  \psi^\mu_{1/2} \tilde\psi^\nu_{1/2} \right)
\\
&&  - \bra{\bar{P}_{(\phi)}} {1 \over 2\sqrt{2}} 
\left(\eta^{\mu\nu} - (\delta^\mu{}_\alpha \delta^\nu{}_\beta +
  \delta^\mu{}_\beta \delta^\nu{}_\alpha)
  k^\alpha \ell^\beta \right) \, ,\nonumber
  \eea
  where now $k^\alpha \eta_{\alpha\beta} \ell^\beta =1$ and
  $\ell^\alpha \ell^\beta \eta_{\alpha\beta}=0$.
  Instead of splitting the projectors in terms of 6-dimensional fields,
  we keep the ten-dimensional notation in order to compare it with the
  non-compact results, and with the approximation made previously.
  As before, the NS-NS fields will have non-zero overlap with the NS-NS
  boundary state. The type of calculation we perform in this case is similar
  to
  the non-compact case, with the difference that there appear
  extra normalisation coefficients, and the momentum dependence is
  only in the spatial directions of the fixed plane.
  Note that the factors of $\prod\limits_{i} \Phi_i^{-1/2}$
  in the normalisation of the boundary state and the massless
  states cancel with a factor $\prod\limits_{i} \Phi_i$ coming from the
  normalisation (\ref{normalisation-kaluza}) in the amplitude.
  We find the following results:
  \bea
  \delta \bar{\phi} (k) &=& {\bra{{\bar P}_{(\phi)}}}
  {\cal D}_{a=1/2} \ket{D0}_{\rm NS,U} =
  { 3 T_0 \over 2\sqrt{2}} \, \prod_i (2\pi R_i)^{-1/2}
  \, {V_1 \over k_\bot^2}
  \, , \nonumber
  \\
  \delta \bar{h}_{\mu\nu} (k) &=& {\bra{{\bar P}_{(h)\, \mu\nu}}}
  {\cal D}_{a=1/2} \ket{D0}_{\rm NS,U}
  \\
  &=& {T_0 \over 2} \, \prod\limits_i (2\pi R_i)^{-1/2}
  \, {V_1 \over k_\bot^2} \,
  {\rm diag} \left( {7\over 4}, {1\over 4}, \dots ,
  {1\over 4} \right) \, .\nonumber
  \eea
  We can make use of the Fourier transformation
  (\ref{Fourier-1}) in the NS-NS sector to derive the spacetime
  behaviour:
  \bea
  \delta \bar{\phi} (y) &=& {T_0 \over 2\sqrt{2} \, \Omega_4}
  \, \prod\limits_i (2\pi R_i)^{-1/2} \, {1 \over |y|^3 } \, ,
  \nonumber \\
  \delta {\bar h}_{\mu\nu} (y) &=& {T_0 \over 6 \, \Omega_4}
  \, \prod\limits_i (2\pi R_i )^{-1/2} \,
  {1 \over |y|^3 } \, {\rm diag}
  \left( {7\over 4}, {1 \over 4} , \dots , {1 \over 4} \right) \, .
  \label{0-mode}
  \eea
  The twisted R-R sector remains the same as for the uncompactified case
  (\ref{RR-asymptotics}).
  Note that the asymptotic behaviour derived
  from the compactified boundary state (\ref{0-mode})
  exactly reproduces the behaviour
  of the 6-dimensional fluctuations
  obtained before (\ref{NSNS-asymptotics-3})
  using the approximation given in (\ref{approximation-2}).
  The asymptotics of the dilaton and the graviton are now certainly comparable
  with that of the twisted R-R 1-form. In particular,
  at the critical radius there is an accidental Bose-Fermi
  degeneracy, which as we will see in the next Section,
  implies a no-force in a brane probe. At the critical radius we can then
  consider a large number of branes $N$, hence the fluctuations take the
  following form:
  \bea
  \delta \bar{\phi}(y) &=& {T_0 N \over 2\sqrt{2} \Omega_4
  (2 \pi^2 \alpha^\prime)} \,
  {1 \over |y|^3} \, ,\nonumber \\
  \delta \bar{h}_{\mu\nu} (y) &=& {T_0 N \over 6 \Omega_4
  (2 \pi^2 \alpha^\prime)} \, {1 \over |y|^3}
  {\rm diag} \left({7 \over 4}, {1 \over 4}, \dots, {1 \over 4} \right)\, ,\\
  \delta C^{(1)} (y) &=& - {Q_0 N \over 3\sqrt{2} \Omega_4} \,
  {1 \over |y|^3} \, .\nonumber
  \label{NSNS-asymptotics-N}
  \eea
  

  \section{No-force Condition at the Critical Radius}


  A composite of branes preserving the BPS property
  satisfies a no-force condition between the constituents.
  This was verified at the level of the effective action and
  background geometries for BPS branes in \cite{Tseytlin}.
  The no-force property is a consequence of the vanishing of the one-loop
amplitude between the D-branes.  
On the other hand, the
  Bose-Fermi degeneracy for non-BPS D-branes at the critical
  radius described in \cite{Gaberdiel-Sen}
  represents a remarkable example in which the no-force property takes place
  without being BPS.
  In this section we recover the no-force condition using the
  non-BPS D-particle as a probe of the geometry of another
  non-BPS D-particle. Firstly, we consider the
  effective action for the non-BPS D-particle.


  \subsection{The Action of the D-particle}


  Following the prescription given in \cite{Sen-action}
  one can construct the action for the D-particle in the orbifold
  of type IIB we are considering. One starts with the action of a
  non-BPS D-particle in type IIB, and set to zero all fields which are odd
  under
  $(-1)^{F_L} {\cal I}_4$. The
  worldvolume theory of this D-particle
  is given by scalars
  $b^\mu_{-1/2}|0{\cal i}_{\rm NS} \otimes \one$,
  $\mu=1, \dots, 9$, a tachyon
  $|0 {\cal i}_{\rm NS} \otimes \sigma_1$,
  and 16 fermions in the Ramond groundstate of each of the Chan-Paton sectors
  $\one$ and $\sigma_1$.
  The orbifold $(-1)^{F_L} {\cal I}_4$ eliminates the
  scalars in the directions $i=6,7,8,9$, the tachyon and
  the fermions with Chan-Paton factor $\sigma_1$
  \cite{Sen-5}. Accordingly, the non-BPS D-particle is stable
  and contains 5 physical bosonic degrees of freedom, describing
  the motion of the D-particle on the orbifold fixed plane, plus
  16 fermionic d.o.f. If we label the coordinates
  as before $X^\mu = (X^\alpha,X^i)$, where $\alpha=0,1,\dots,5$
  labels the orbifold
  fixed plane directions, we can write the following action for the
  D-particle:
  \be
  S_{kin}= - {T_{0} \over \kappa_{10}} \int d\tau \,
  {\rm e}^{-\phi} \sqrt{|\partial_\tau X^\alpha \partial_\tau X^\beta
  \, G_{\alpha\beta} |} \, ,
  \label{kinetic}
  \ee
  where $G_{\mu \nu}$ is the background string metric.
  Furthermore, since the D-particle carries charge under the twisted
  R-R 1-form ${\cal C}^{(1)}$, we can include
  a WZ coupling\footnote{Wess-Zumino couplings for
  stable non-BPS D-branes have been recently considered in
  \cite{Russo-Scrucca}.}, which
  involves the 1-form potential on the fixed plane:
  \be
  S_{WZ} = {{Q}_{0} \over 2\kappa_{6}} \int \partial_\tau X^\alpha
  {\cal C}^{(1)}_\alpha \, .
  \label{WZ}
  \ee
  The couplings of the stable non-BPS D-particle
  to the background fields given by equations (\ref{kinetic}) and (\ref{WZ})
  can be also derived by projecting the boundary state of the D-particle
  onto the projectors defined previously, analogously
  to the BPS case considered in \cite{Divecchia-1}. The coupling to the
  dilaton is given by
  \be
  {\cal J}_{(\varphi)} =
  \bra{P_{(\phi)}} D0 \rangle_{\rm NS,U} = {3 T_0 \over 2\sqrt{2}} V_1 \, ,
  \ee
  whereas the coupling to the metric is given by
  \be
  {\cal J}_{(h)} =
  \eta_{\mu\nu}
  \bra{P_{(h)}^{\mu\nu}} D0 \rangle_{\rm NS,U} = {T_0 V_1 h_{00}} \, ,
  \ee
  where $h_{\mu\nu}$ is the symmetric and traceless
  helicity tensor of the graviton. These two results reproduce the couplings
  of (\ref{kinetic}) in Einstein frame ($g_{\mu\nu} =
  {\rm e}^{-\phi/2} G_{\mu\nu}$) after rescaling the dilaton
  as $\phi = \kappa_{10} \sqrt{2} \varphi$.
  Finally, the coupling to the twisted R-R 1-form is given by:
  \be
  {\cal J}_{(C)} = C^{(1)}_\alpha \,
  \bra{P_{(C)}^{\alpha}} D0 \rangle_{\rm R,T} =
  {Q_0 \over \sqrt{2}} V_1 C^{(1)}_0 \, ,
  \ee
  which reproduces (\ref{WZ}), after the rescaling
  ${\cal C}^{(1)} = \kappa_{6} \sqrt{2} C^{(1)}$.

  In the compact case, the coupling of the D-particle to the massless closed
  string fields can be obtained as above, this time
  using the compactified boundary state
  constructed in Section \ref{compact-boundary}. The net result is a
  change of the factor in front of the kinetic term, which amounts
  to the relation of the 6-dimensional
  gravitational constant with the 10-dimensional one. The
  boundary state reproduces the couplings of the
  following action:
  \be
  S_{kin}= - {T_{0} \over \kappa_6}
  \prod\limits_{i=6}^9 (2\pi R_i)^{-1/2}
  \int d\tau \,
  {\rm e}^{-\phi} \sqrt{|\partial_\tau X^\alpha \partial_\tau X^\beta
  \, G_{\alpha\beta} |}
  \, .
  \ee
  Thus although the D-particle does not couple to the components
  $G_{ij}$ (\ref{kinetic}), it does feel the transverse geometry through the
  renormalised tension.
  Moreover, the open strings on the D-particle
  carry winding modes, and at the critical radius, the
  winding modes of the tachyon become massless.
  These extra modes appear as new degrees of freedom, $\chi^i$,
  in the effective action \cite{Majumder-Sen,Sen-action}:
  \be
  S_{kin}= - {T_{0} \over 2\pi^2 \alpha^\prime \kappa_6} \int d\tau \,
  {\rm e}^{-\phi} \sqrt{|\partial_\tau X^\alpha \partial_\tau X^\beta
  \, G_{\alpha\beta} |}
  \left( 1 - G^{\tau\tau} \partial_\tau \chi^i \partial_\tau \chi^i \right)
  \, .
  \label{kinetic-critical}
  \ee
  However, these extra modes will not play any role in our
  discussion below.

  \subsection{D-particle at the Critical Radius}

  We consider a non-BPS D-particle probe moving in the background
  of another non-BPS D-particle in the compactified orbifold.
  We split the embedding coordinates as $X^\alpha=(X^0,X^m)$,
  $m=1,\dots,5$, and choose the static gauge:
  \be
  X^0 = \tau \, .
  \ee
  We identify the embedding scalars with the transverse coordinates of the
  background geometry and expand the action in derivatives\footnote{We 
  use the Einstein frame
  since we want to use in the action the results given by the boundary state
  computation.}:
  \be
  \sqrt{|\partial_\tau y^\alpha \partial_\tau y^\beta g_{\alpha\beta}|}
  = \sqrt{|g_{00}|} \left( 1 - {1 \over 2 g_{00}} \partial_0 y^m
  \partial_0 y^n g_{mn} + \dots \right) \, .
  \ee
  Moreover, for a D-particle background, only the zeroth
  component of the twisted R-R 1-form is turned on:
  \be
  \partial_\tau X^\alpha {\cal C}^{(1)}_\alpha = {\cal C}^{(1)}_0 \, .
  \ee
  The action can then be approximated by
  \be
  S \simeq \int d\tau \, {\cal V}(y)
  + {{T}_{0} \over \kappa_6} \prod\limits_{i=6}^9
  (2\pi R_i)^{-1/2} \int d\tau \, {{\rm e}^{-{3 \over 4}\phi}
  \over 2 \sqrt{|g_{00}|}}
  \partial_0 y^m
  \partial_0 y^n g_{mn} \, ,
  \label{action-approx}
  \ee
  where we have defined the effective static potential ${\cal V}$:
  \be
  {\cal V} (y) = - {{T}_{0} \over \kappa_6}
  \prod\limits_{i=6}^9 (2\pi R_i)^{-1/2}
  \, {\rm e}^{-{3 \over 4}\phi} \, \sqrt{|g_{00}|}
  \, + \,
  {{Q}_{0} \over {2} \kappa_6}\, {\cal C}^{(1)}_0 \, .
  \label{effective-static-potential}
  \ee
  If this potential is not constant or zero for a given background
  $g_{00}(y)$, $\phi(y)$, ${\cal C}^{(1)}_0(y)$, there will be a force term
  in the field equations for $y^m$. Plugging into this potential the
  background
  geometry generated by
  another D-particle, one can check whether
  the Bose-Fermi degeneracy takes place.
  Since from the boundary state we obtain
  the asymptotic form of the D-particle solution,
  we rewrite the potential ${\cal V}$
  in terms of the fluctuations
  (\ref{generic-fluctuation-6-dim})\footnote{We
  neglect the constant parts of the potential, since these would not generate
  any force.}:
  \be
  {\cal V} (y) \simeq {T_0}
  \prod\limits_{i=6}^9 (2\pi R_i)^{-1/2}
  \left( {3 \over 2\sqrt{2}} \delta \bar{\phi} + \delta \bar{h}_{00} \right)
  + {Q_0 \over \sqrt{2}} \delta C^{(1)}_0 \, .
  \ee
  We can insert now the asymptotic form of the solution
  for the compactified case given by
  (\ref{0-mode}) and (\ref{RR-asymptotics}).
  The potential then takes the form:
  \bea
  {\cal V} (y) &=& {2 \over 3 \, \Omega_4 \, |y|^3}
  \left( T_0^2 \prod_{i=6}^9 (2\pi R_i)^{-1} - {Q_0^2 \over 4}\right)
  \nonumber \\
  &=& {4 \pi \alpha^\prime \over |y|^3}
  \left( \left({\alpha^\prime \over 2}\right)^2
  \prod_{i=6}^9 R_i^{-1} - 1 \right) \, ,
  \eea
  which coincides with the static potential derived from the cylinder amplitude
in (\ref{scattering-potential}) 
and vanishes for the critical radius ($R_i = \sqrt{\alpha^\prime/2}$),
  hence we recover the no-force at the critical radius using the background
  geometry generated by the non-BPS D-particles.

 According to the interaction potential
${\cal U}_{open}$ derived in Section 2.3,
the velocity corrections start at order $v^4$ and 
there is no $v^2$ corrections to the
potential in for any radius, as happens for 
BPS branes. From the probe point of view, this translates
into the fact that the
{\it metric} on the moduli space, multiplying the velocity dependent piece
of the action (\ref{action-approx}), is flat \cite{Tseytlin}. 
In the asymptotic limit, for the non-compact orbifold this metric
takes the form:
  \be
  {e^{-{3\over 4}\phi } \over \sqrt{|g_{00}|}}
  g_{mn} \simeq
  \delta_{mn} \left(1-{3 \over 2\sqrt{2}} \kappa_{10}
  \delta {\phi}(x) + \kappa_{10} \delta  {h}_{00}(x) \right)
  +2 \kappa_{10} \delta {h}_{mn} (x) \, .
  \label{velocity-probe}
  \ee
Using the asymptotic fluctuations
given in (\ref{NSNS-asymptotics}), it is easy
to see that the metric is flat ($=\delta_{mn}$).
In the case of the compact orbifold, this metric factor takes the form:
\be
  {e^{-{3\over 4}\phi } \over \sqrt{|g_{00}|}}
  g_{mn} \simeq
  \delta_{mn} \left(1-{3 \over 2\sqrt{2}} \kappa_{6}
  \delta \bar{\phi}(y) + \kappa_{6} \delta \bar{h}_{00}(y) \right)
  +2 \kappa_{6} \delta \bar{h}_{mn} (y) \, .
  \label{velocity-probe-2}
\ee
Substituting the asymptotic fluctuations (\ref{0-mode}),
  we obtain the same flat metric for any value of the radii.
Thus we recover the behaviour given by the open strings in
(\ref{potential-non-compact-open}) and (\ref{scattering-potential-open}).

  \subsection{The Classical Geometry of the non-BPS D-particle}

  In this Section we assume that the 
  no-force condition will persist at the full
  non-linear level of the field equations.
  We can then restrict the possible classical
  geometries by imposing the no-force
  at the level of equation (\ref{effective-static-potential}).
  No-force can occur for ${\cal V}$ either constant or zero
  \cite{Tseytlin}. However, in our case, the no-force condition at the
  critical radius must enforce the relation
  $T_0 = (\pi^2 \alpha^\prime) Q_0$, as seen in (\ref{bose-fermi-deg}).
  These are in fact the factors of the NS-NS and R-R contributions of the
  potential (\ref{effective-static-potential}), respectively.
  Hence we conclude that
  the classical solution must be such that at the critical radius
  the following equality holds
  \be
  e^{-{3 \over 4}\phi} \, \sqrt{|g_{00}|} = {\cal C}^{(1)}_0 \, .
  \label{critical-equality}
  \ee
  We can then deduce part of the form for
  $g_{00}$, $\phi$, and ${\cal C}^{(1)}$:
  \bea
  g_{00} (y) &=& - \, \left( 1  + {\kappa_6 T_0 \over 2a \Omega_4}
  (2\pi^2 \alpha^\prime)^{-1} \, {1 \over |y|^3}
  + \dots \right)^{-{7\over 6}a}
  \, ,\nonumber \\
  {\rm e}^\phi (y) &=& \left( 1  + {\kappa_6 T_0 \over 2a \Omega_4}
  (2\pi \alpha^\prime)^{-1} \, {1 \over |y|^3}
  + \dots \right)^a
  \, ,
  \label{assumption}
  \\
  {\cal C}^{(1)}_0 (y) &=& \left( 1  + {\kappa_6 Q_0 \over 4 a
  \Omega_4}
  \, {1 \over |y|^3} + \dots \right)^{-{4\over 3}a} -1 \, .\nonumber
  \eea
  These functions are given in terms of a parameter $a$ yet to be determined,
  and are such that asymptotically they become
  the fluctuations in (\ref{NSNS-asymptotics-3})
  and (\ref{RR-asymptotics}), and at the critical radius
  equation (\ref{critical-equality}) holds.
  The dots indicate other possible contributions with dependence $|y|^n$,
  $n < -3$, which are subleading in the asymptotic limit
  ($|y| \to \infty$), but which are relevant when we come closer to the
  brane.
  We expect these extra terms to appear
  since harmonicity,
  which is a direct consequence of supersymmetry,
  may not be present in the complete solution.
  
We can make one further assumption if we consider that
  the metric factor in the velocity dependent piece of the action
(\ref{action-approx})
  remains flat for the complete geometry.
  We impose then the following relation:
  \be
  {{\rm e}^{-{3\over 4}\phi } \over \sqrt{|g_{00}|}}
  g_{mn} = \delta_{mn} \, .
  \label{velocity-metric}
  \ee
  Note that the functions describing
  the dilaton and the metric must be the same
  at the critical radius. Moreover, it is only at this radius
  where we actually expect to find a consistent classical geometry.
  Therefore, we make use of the expressions
  for $g_{00}(y)$ and $\phi(y)$ given above to obtain:
  \be
  g_{mn} (y) = \left( 1 + {\kappa_6 T_0 \over 2a \Omega_4}
  (2\pi^2 \alpha^\prime)^{-1} \, {1 \over |y|^3} + \dots
  \right)^{{1\over 6}a} \delta_{mn} \, ,
  \ee
  where the dots represent the same contributions as in
  (\ref{assumption}) at the critical radius.
  Finally, it seems that $g_{ij}$ is out of the reach of the present analysis.
A possible way to include it in the analysis is 
by a coupling to the extra 
massless states $\chi^i$ in (\ref{kinetic-critical}). 
On the other hand, although the asymptotic behaviour of the metric
  in the non-compact case
  (\ref{NSNS-asymptotics}) presents an $SO(9)$ symmetry,
  this is certainly broken to $SO(5) \times SO(4)$ by the orbifold
  when we get closer to the position of the brane. This is
  in fact already suggested by
  the asymptotics in the compact case (\ref{NSNS-asymptotics-3}).
  Accordingly, the form of $g_{ij}(y)$
  is expected to be different from $g_{mn}(y)$.


  \section{Comments}


  In this paper we have investigated the description of a stable
  non-BPS D-particle in terms of a classical solution.
  We have used the technique of the boundary state to obtain the
  asymptotic form of the classical solution for the
  non-compact and compact versions of the orbifold.
  We find a metric and a dilaton propagating in the bulk, and
  a twisted R-R 1-form propagating in the fixed plane.
  In the non-compact case the bulk fields have the dependence expected for
  a particle in ten dimensions, whereas in the compact case they
  have a dependence
  typical of a particle in six dimensions.
  The twisted R-R 1-form has in both cases the same asymptotic
  form with the usual dependence for a particle in six-dimensions.
  Using the non-BPS D-particle as a probe in the background of another
  non-BPS D-particle, we have recovered the no-force
  property at the critical radius using the asymptotic behaviour.
  Moreover, we have calculated the cylinder amplitude for non-BPS 
D-particles in relative motion. From it we have extracted the
long and short distance interactions. For generic radii these contain
$v^2$ corrections in the closed string description,
but the open string description presents no $v^2$ terms
for all radii, like for BPS D-branes. Moreover, 
the $v^4$ corrections do not match, 
unlike for BPS branes. On the other hand,
at the critical radius they present a BPS-like behaviour, up to the $v^4$ 
corrections, which do not match in the open and closed descriptions.

We have assumed that the no-force
  property of a brane probe holds for the full background geometry. 
This is acceptable for distances
much larger than the string scale.
  This assumption allows us 
to derive part of the classical solution, which reproduces
  the asymptotic behaviour and the no-force property.
  On the other hand, we expect that extra terms may appear
  in the solution at the classical level.
  This is due to the fact that
  the boundary state only gives information
  about the next-to-leading term in the asymptotic limit, hence
  subleading terms 
  which become relevant at short distances
  escape from this analysis.
  Moreover, the fact that there is a coordinate system in which the
  metric can be written in terms of harmonic functions is very much
  related to residual supersymmetry, which does not occur
  in our case.
  This fact does not permit to find information about the geometry
  near the core of the non-BPS D-particle.
  Moreover, although the asymptotic behaviour of the metric
  (\ref{NSNS-asymptotics}) presents an $SO(9)$ symmetry,
  this is expected to be broken
  to $SO(5) \times SO(4)$ by the orbifold
  when we get closer to the position of the brane.

  The form of the asymptotic behaviour
  we have found suggests that the classical solution
  for the non-BPS D-particle will be a solution
  to the field equations derived
  from an action involving 10-dimensional bulk fields
  and 6-dimensional matter fields constrained on an orbifold fixed plane:
  \be
  {\cal S}_{total} = {\cal S}_{bulk} + {\cal S}_{plane} \, .
  \ee
  The bulk action involves the metric and the dilaton, and other fields
  of the massless untwisted sector. For the case of the D-particle,
  only the metric and dilaton are relevant, hence in Einstein frame
  we have:
  \be
  {\cal S}_{bulk} = {1 \over 2\kappa^2_{10}}
  \int d^{10} x \, \sqrt{|{\rm det}(g_{\mu\nu})|}
  \left( {\cal R} - {1 \over 2} (\partial \phi)^2 \right) \, .
  \ee
  Since the twisted sector is provided by a type IIB NS-5 brane
  hidden in the orbifold fixed plane, we can derive
  the fixed plane action from the effective action of the type IIB
  NS-5 brane in Einstein frame:
  \be
  {\cal S}_{plane}
  = m \int\limits_{x^i=0} d^6 y \sqrt{|{\rm det}({\tilde g}_{\alpha\beta})|}
  {{\rm e}^{- {3 \over 2}\tilde{\phi}} \over 4}
  F_{\alpha\beta} F^{\alpha\beta} \, ,
  \ee
  where the tilde denotes the restriction of the bulk fields to the
  position of the fixed plane:
  \be
  {\tilde g}_{\alpha\beta} (y^\alpha)
  = g_{\alpha\beta} (y^\alpha,x^i=0) \, ,\qquad
  {\tilde \phi}(y^\alpha) = \phi (y^\alpha, x^i=0) \, ,
  \ee
  and $\alpha,\beta =0,1,\dots,5$
  are the indices along the fixed plane.
  This particular coupling of the twisted fields with the bulk metric is
  obtained by expanding the NS-5 brane kinetic term
  in powers of the worldvolume fields.
  The twisted R-R 1-form has been identified with the
  $U(1)$ gauge field on the NS-5 brane worldvolume $F$.
  Here $m$ is a factor related to the tension of the
  NS-5 brane.
  Moreover, the embedding
  scalars of the NS-5 brane has not been included,
  since they correspond to the twisted NS-NS sector,
  to which the non-BPS D-particle does not couple.
  It is straightforward to see that the asymptotic fields
  (\ref{NSNS-asymptotics}) and (\ref{RR-asymptotics})
  are a solution to the weak field limit of the equations of motion
  of the action ${\cal S}_{total}$. Note that the non-BPS D-particle
is not stable below the critical radius, therefore we do not expect
it would appear as a solution of this action reduced to six dimensions.

Finally, the behaviour of the stable non-BPS D-particle at the critical radius
suggests that it probably saturates a BPS type of bound in the effective 
theory, without being supersymmetric. This latter property comes 
from the fact that
the particle couples to a 1-form in the {\it wrong} supersymmetric
multiplet. The analysis carried out here can be extended to
other stable non-BPS branes.  
A study of the classical geometry 
of other stable non-BPS D-branes will be presented in a future
publication \cite{preparation}.


  \section*{Acknowledgements}

  We would like to thank E.~Bergshoeff and M.~de Roo
  for useful discussions.
  E.E. is greatful to M.~Gaberdiel, T.~Dasgupta, N.~Lambert, A.~Liccardo,
  M.J.~Perry, B.~Stefanski, P.~Townsend and A.~Uranga for useful discussions.
  E.E. has also enjoyed discussions with
  C.~van der Bruck, P.~Vanhove and N.~Wyllard.
  The work of E.E. is supported by the European Community
  program "Human Potential"
  under the contract HPMF-CT-1999-00018.
  This work is also partially supported by
  the PPARC grant PPA/G/S/1998/00613.


  \appendix
  \section{Appendix}
  \label{appendixA}

  \noindent In this Appendix we present some details about the
  zero-mode part of the
  twisted R-R boundary state and its GSO projection.
  In order to describe the twisted R-R groundstate we make use of the
  $8\times 8$ gamma matrices of $SO(1,5)$:
  \be
  \lbrace \gamma^\alpha , \gamma^\beta \rbrace = 2 \one_8 \,
  \eta^{\alpha \beta} \, .
  \ee
  We define
  \be
  \gamma = - \gamma^0 \gamma^1 \gamma^2 \gamma^3 \gamma^4 \gamma^5 \, ,
  \ee
  such that $\lbrace \gamma^\alpha, \gamma \rbrace =0$
  and $(\gamma)^2= \one_8$.
  Furthermore, there is a conjugation matrix
  \be
  {\cal C} = \gamma^3 \gamma^5 \gamma^0 \, ,
  \ee
  such that
  \be
  \lbrace \gamma , {\cal C} \rbrace = 0
  \, ,\qquad
  (\gamma^\alpha)^T = - {\cal C} \gamma^\alpha {\cal C}^{-1}
  \, ,\qquad {\cal C}^{-1} = {\cal C} \, ,\qquad {\cal C}^T = {\cal C} \, .
  \ee
  The twisted R-R groundstate is characterised by
  left and right spinor indices of $SO(1,5)$, and can be constructed
  from the NS-vacuum by means of spin and twist fields as follows:
  \be
  \lim\limits_{z,\bar{z} \to 0}
  S^a (z) \Sigma(z) {\tilde S}^b(\bar{z}) {\tilde \Sigma}(\bar{z})\ket{0}
  = \ket{a}_{\rm T} \widetilde{\ket{b}}_{\rm T} \, .
  \ee
  The zero-mode part of the twisted R-R boundary state satisfies the
  following overlap equations in the twisted sector:
  \be
  \left(\psi_0^{\alpha} - {\rm i}
  \eta \, S^\alpha{}_\beta \, \tilde{\psi}_0^\beta \right)
  \ket{D0_{\psi}, \eta}_{\rm R,T}^{(0)} = 0 \, ,
  \ee
  where $\alpha = 0, 1, \dots 5$ and $S^\alpha{}_\beta$ is the restriction
  of the matrix $S^\mu{}_\nu$ of the D-particle
  to the 6-dimensional orbifold fixed plane:
  $S^\alpha{}_\beta = {\rm diag}(+1,-1,-1,-1,-1,-1)$. We define
  \be
  \ket{D0_{\psi}, \eta}_{\rm R,T}^{(0)} = M_{ab} \ket{a}_{\rm T}
  \widetilde{\ket{b}}_{\rm T} \, .
  \ee
  If we define the action of the fermionic zero-modes in the twisted
  Ramond sector as
  \begin{eqnarray}
  \psi_0^\alpha \ket{a}_{\rm T} \widetilde{\ket{b}}_{\rm T} &=& {1 \over
  \sqrt2}
  (\gamma^\alpha)^a{}_c (\one_8)^b{}_d \, \ket{c}_{\rm T}
  \widetilde{\ket{d}}_{\rm T} \, ,
  \nonumber \\
  \tilde\psi_0^\beta \ket{a}_{\rm T} \widetilde{\ket{b}}_{\rm T}
  &=& {1 \over \sqrt2}
  (\gamma)^a{}_c (\gamma^\beta)^b{}_d \, \ket{c}_{\rm T}
  \widetilde{\ket{d}}_{\rm T} \, ,
  \label{zeromode-1}
  \end{eqnarray}
  the matrix $M$ must satisfy:
  \be
  (\gamma^\alpha)^T M- {\rm i}\eta \, S^\alpha{}_\beta
  \left( \gamma M \gamma^\beta \right) = 0 \, .
  \ee
  The solution to this equation is given by:
  \be
  M= {\cal C} \gamma^0 \, {1 + i \eta \gamma \over 1 + i\eta} \, .
  \label{M-matrix}
  \ee
  The zero-mode R-R boundary state for a non-BPS D-particle  moving in
a direction $m$ is given by
\be
  \ket{D0_{\psi}, \eta, v}_{\rm R,T}^{(0)} = M_{ab}(v) \ket{a}_{\rm T}
  \widetilde{\ket{b}}_{\rm T} \, ,
  \ee
where the matrix $M_{ab}(v)$ is given by
\be
M (v) = {1 \over \sqrt{1 - v^2}} {\cal C} 
\left(\gamma^0 +v \gamma^m \right) {1 + i\eta \gamma \over 1 + i\eta}
\label{M-boosted}
\ee  
On the other hand, defining
  \be
  {}_{\rm R,T}^{(0)}\bra{D0_{\psi}, \eta}
  = {}_{\rm T}\bra{a} {}_{\rm T}\widetilde{\bra{b}} N_{ab} \, .
  \ee
  and using
  \begin{eqnarray}
  {}_{\rm T}\bra{a} {}_{\rm T}\widetilde{\bra{b}}\, \psi_0^\alpha
  &=& - {1 \over \sqrt2}
  {}_{\rm T}\bra{c}
  {}_{\rm T}\widetilde{\bra{d}}
  \, (\gamma^\alpha)^a{}_c (\one_8)^b{}_d \, \nonumber \\
  {}_{\rm T}\bra{a}
  {}_{\rm T}\widetilde{\bra{b}}
  \, \tilde\psi_0^\beta
  &=& {1 \over \sqrt2}
  {}_{\rm T}\bra{c}
  {}_{\rm T}\widetilde{\bra{d}}
  \, (\gamma)^a{}_c (\gamma^\beta)^b{}_d \,
  \label{zeromode-2}
  \end{eqnarray}
  we can write the matrix $N$ as follows:
  \be
  N = {\cal C} \gamma^0 {1 + i \eta \gamma \over 1 - i \eta} \, .
  \ee
The boosted version can be written similarly to (\ref{M-boosted}):
\be
 N(v) = {{1 \over \sqrt{1 - v^2}} 
\cal C} \left(\gamma^0 +v \gamma^m \right)
{1 + i \eta \gamma \over 1 - i \eta} \, .
\ee
  In order to implement the
  GSO projection on the zero-mode of the boundary state,
  we define the zero-mode part of the GSO-projector on the twisted
  R-R sector as:
  \be
  \Psi = - 8 \psi_0^5 \cdots \psi_0^0 \, ,\qquad
  \tilde\Psi = - 8 \tilde\psi_0^5 \cdots \tilde\psi_0^0 \, .
  \ee
  Using this definition and (\ref{zeromode-1}),
  the action of $\Psi$ and $\tilde\Psi$
  on the twisted R-R groundstate is found to be:
  \be
  \Psi \ket{a}_{\rm T} \widetilde{\ket{b}}_{\rm T} =
  (\gamma)^a{}_c \, \ket{c}_{\rm T} \widetilde{\ket{b}}_{\rm T}
  \, ,\qquad
  \tilde\Psi \ket{a}_{\rm T} \widetilde{\ket{b}}_{\rm T} =
  (\gamma)^b{}_d \, \ket{a}_{\rm T} \widetilde{\ket{d}}_{\rm T}
  \, .
  \label{zero-mode-GSO-projector}
  \ee
  Finally, combining (\ref{M-matrix})
  and (\ref{zero-mode-GSO-projector}) we find:
  \be
  \Psi \ket{D0_{\psi}, \eta}_{\rm R,T}^{(0)}
  = \ket{D0_{\psi}, - \eta}_{\rm R,T}^{(0)}
  \, ,\qquad
  \tilde\Psi \ket{D0_{\psi}, \eta}_{\rm R,T}^{(0)}
  = \ket{D0_{\psi}, - \eta}_{\rm R,T}^{(0)} \, .
  \ee


  \section{Appendix}
  \label{appendixB}

  For the sake of completeness we include in this Appendix the
  explicit definitions of the different parts of the boundary state for the
  non-BPS D-particle. We include as well the conjugate states.
  Before GSO-projection the NS-NS boundary state is given by:
  \be
  \ket{D0,\eta}_{\rm NS,U} = {T_0 \over 2}
  \ket{D0_X}\, \ket{D0_{\rm gh}}\,{\ket{D0_\psi,\eta}}_{\rm NS}
  \,{\ket{D0_{\rm sgh},\eta}}_{\rm NS} \, .
  \ee
  The bosonic part is:
  \be
  \ket{D0_X} =
  \delta^{(5)}({\hat q}^p-y^p) \delta^{(4)}({\hat q}^i)
  \, {\rm exp}\left( -\sum\limits_{n=1}^\infty {1 \over n}
  \alpha_{-n} \cdot S \cdot \tilde{\alpha}_{-n} \right) \ket{k=0}
  \, .
  \ee
  The ghost part:
  \be
  \ket{D0_{\rm gh}} =
  {\rm exp}\left( \sum\limits_{n=1}^\infty
  (c_{-n} {\tilde b}_{-n} - b_{-n} {\tilde c}_{-n} ) \right)
  \left( {c_0 + \tilde{c}_0 \over 2}\right) \ket{1}\widetilde{\ket{1}}
  \, .
  \ee
  The fermionic part:
  \be
  \ket{D0_\psi, \eta}_{\rm NS} =
  -i \, {\rm exp}\left( i\eta \sum\limits_{r=1/2}^\infty
  \psi_{-r} \cdot S \cdot \tilde{\psi}_{-r} \right) \ket{0} \, ,
  \ee
  The superghost part:
  \be
  \ket{D0_{\rm sgh}, \eta}_{\rm NS} =
  {\rm exp}\left( i\eta \sum\limits_{r=1/2}^\infty
  (\gamma_{-r} \tilde{\beta}_{-r} - \beta_{-r} \tilde{\gamma}_{-r}) \right)
  \ket{-1} \widetilde{\ket{-1}}
  \, .
  \ee
  The conjugate states are:
  \bea
  \bra{D0_X} &=&
  \bra{k=0} \delta^{(5)}({\hat q}^p- y^p) \delta^{(4)}({\hat q}^i)
  \, {\rm exp}\left( -\sum\limits_{n=1}^\infty {1 \over n}
  \alpha_{n} \cdot S \cdot \tilde{\alpha}_{n} \right) \, ,
  \nonumber \\
  \bra{D0_{\rm gh}} &=&
  \bra{2} \widetilde{\bra{2}}
  \left(b_0 - \tilde{b}_0\right)
  {\rm exp}\left( \sum\limits_{n=1}^\infty
  ({\tilde b}_{n} c_n - {\tilde c}_{n} b_n) \right)
  \, ,\nonumber \\
  {}_{\rm NS}\bra{D0_\psi, \eta} &=&
  i \, \bra{0} {\rm exp}\left( i\eta \sum\limits_{r=1/2}^\infty
  \psi_{r} \cdot S \cdot \tilde{\psi}_{r} \right) \, ,
  \nonumber \\
  {}_{\rm NS}\bra{D0_{\rm sgh}, \eta} &=&
  \bra{-1}\widetilde{\bra{-1}}
  \, {\rm exp}\left( - i\eta \sum\limits_{r=1/2}^\infty
  ({\beta}_{r} \tilde{\gamma}_r - \gamma_r \tilde{\beta}_{r}) \right) \, .
  \eea
  For the twisted R-R sector we have (before GSO-projection):
  \be
  \ket{D0,\eta}_{\rm R,T} = {Q_0 \over 2}
  \ket{D0_X}_{\rm T} \, \ket{D0_{\rm gh}} \,
  \ket{D0_\psi, \eta}_{\rm R,T} \, \ket{D0_{\rm sgh}, \eta}_{\rm R} \, ,
  \ee

  \noindent The bosonic part is given by
  \be
  \ket{D0_X}_{\rm T} =
  \delta^{(5)}({\hat q}^p-y^p) \,
  \, {\rm exp}\left( -\sum\limits_{t.m.} {1 \over n}
  \alpha_{-n} \cdot S \cdot \tilde{\alpha}_{-n} \right) \ket{k=0}
  \ee
  where $t.m.$ indicates that the sum is performed according to the
  {\it twisted moddings} of the (twisted) R-R sector given
  in (\ref{twisted-modding}).
  The ghost part is the same as for the NS-NS boundary state.
  The fermionic part reads:
  \be
  \ket{D0_\psi, \eta}_{\rm R,T} =
  - {\rm exp}\left( i\eta \sum\limits_{t.m.}
  \psi_{-n} \cdot S \cdot \tilde{\psi}_{-n} \right)
  \ket{D0_\psi, \eta}^{(0)}_{\rm R,T} \, ,
  \ee
  where $\ket{D0_\psi, \eta}^{(0)}_{\rm R,T}$
  and is given in Appendix
  \ref{appendixA}.
  Finally, the superghost part:
  \be
  \ket{D0_{\rm sgh}, \eta}_{\rm R} =
  {\rm exp}\left( i\eta \sum\limits_{n=1}^\infty
  (\gamma_{-n} \tilde{\beta}_{-n} - \beta_{-n} \tilde{\gamma}_{-n}) \right)
  \ket{D0_{\rm sgh}, \eta}^{(0)}_{\rm R}
  \, ,
  \ee
  with the zero-mode given by
  \be
  \ket{D0_{\rm sgh}, \eta}^{(0)}_{\rm R}
  = {\rm e}^{i\eta \gamma_0 \tilde{\beta}_0}
  \ket{-1/2} \widetilde{\ket{-3/2}} \, .
  \ee
  Finally, the conjugate states are:
  \bea
  {}_{\rm T}\bra{D0_X} &=&
  \bra{k=0} \delta^{(5)}({\hat q}^p-y^p)
  \, {\rm exp}\left( -\sum\limits_{t.m.} {1 \over n}
  \alpha_{n} \cdot S \cdot \tilde{\alpha}_{n} \right) \, ,
  \nonumber \\
  {}_{\rm R,T}\bra{D0_\psi, \eta} &=& - \,
  {}^{(0)}_{\rm R,T}\bra{D0_\psi, \eta}
  {\rm exp}\left( i\eta \sum\limits_{t.m.}
  \psi_{n} \cdot S \cdot \tilde{\psi}_{n} \right) \, ,
  \nonumber \\
  {}_{\rm NS}\bra{D0_{\rm sgh}, \eta} &=&
  {}^{(0)}_{\rm R}\bra{D0_{\rm sgh}, \eta}
  \, {\rm exp}\left( - i\eta \sum\limits_{n=1}^\infty
  ({\beta}_{n} \tilde{\gamma}_n - \gamma_n \tilde{\beta}_{n}) \right) \, ,
  \eea
  where ${}^{(0)}_{\rm R} \bra{D0_{\psi}, \eta}$
  is given in Appendix \ref{appendixA} and
  \be
  {}^{(0)}_{\rm R}\bra{D0_{\rm sgh}, \eta}
  = \bra{-3/2} \widetilde{\bra{-1/2}}
  \, {\rm e}^{- i\eta \beta_0 \tilde{\gamma}_0} \, .
  \ee


  \end{document}